\documentclass[]{aa}
\usepackage{graphicx}
\usepackage[varg]{txfonts}
\usepackage{natbib}
\usepackage[fleqn]{amsmath}
\usepackage[modulo, switch]{lineno}
\usepackage{nicefrac}
\usepackage{textcomp}
\usepackage{xfrac}
\usepackage{hyperref}

\usepackage{txfonts}
\def\mh{\,$\mu$Hz}
\def\teff{$T_{\mathrm{eff}}$}
\def\lg{\ensuremath{\log g}}
\def\den{$\bar\rho$}
\def\num{$\nu_\mathrm{max}$}
\def\dnu{$\Delta\nu$}
\def\dc{$\Delta\nu_\mathrm{c}$}
\def\dcor{$\Delta\nu_\mathrm{cor}$}
\def\sun{\hbox{$_\odot$}}
\def\kep{\textit{Kepler}}
\def\na{NGC\,6791}
\def\nb{NGC\,6819}

\begin{document}
%
   \title{Nonlinear seismic scaling relations}

\author{T. Kallinger\inst{1}
\and	P.\,G. Beck\inst{2,3}
\and D. Stello\inst{4,5,6}
\and R. A. Garcia\inst{7,8}
            }

   \offprints{thomas.kallinger@univie.ac.at}

   \institute{
Institute for Astrophysics, University of Vienna, T\"urkenschanzstrasse 17, 1180 Vienna, Austria
              	\and
Instituto de Astrof\'{\i}sica de Canarias, E-38200 La Laguna, Tenerife, Spain
	\and
Departamento de Astrof\'{\i}sica, Universidad de La Laguna, E-38206 La Laguna, Tenerife, Spain
	\and
School of Physics, University of New South Wales, NSW 2052, Australia
	\and
Sydney Institute for Astronomy (SIfA), School of Physics, University of Sydney, NSW 2006, Australia
	\and
Stellar Astrophysics Centre, Department of Physics and Astronomy, Aarhus University, Ny Munkegade 120, DK-8000 Aarhus C, Denmark
	\and
IRFU, CEA, Universit\'e Paris-Saclay, F-91191 Gif-sur-Yvette, France
	\and
Universit\'e Paris Diderot, AIM, Sorbonne Paris Cit\'e, CEA, CNRS, F-91191 Gif-sur-Yvette, France
}

   \date{Received 14 February 2018 / Accepted 16 May 2018}

\abstract
{In recent years the global seismic scaling relations for the frequency of maximum power, $\nu_\mathrm{max} \propto g/\sqrt{T_\mathrm{eff}}$, and for the large frequency separation, $\Delta\nu \propto \sqrt{\bar\rho}$, have caught the attention of various fields of astrophysics. This is because these relations can be used to estimate parameters, such as mass and radius of stars that show solar-like oscillations. With the exquisite photometry of \textit{Kepler}, the uncertainties in the seismic observables are small enough to estimate masses and radii with a precision of only a few per cent. Even though this seems to work quite well for main-sequence stars, there is empirical evidence, mainly from  studies of eclipsing binary systems, that the seismic scaling relations systematically overestimate the mass and radius of red giants by about 15 and 5\%, respectively. Various model-based corrections of the $\Delta\nu -$scaling reduce the problem but do not solve it.}
{Our goal is to define revised seismic scaling relations that account for the known systematic mass and radius discrepancies in a completely model-independent way.}
{We use probabilistic methods to analyse the seismic data and to derive nonlinear scaling relations based on a sample of six red-giant branch (RGB) stars that are members of eclipsing binary systems and about 60 red giants on the RGB as well as in the core-helium burning red clump (RC) in the two open clusters \na\ and \nb .}
{We re-examine the global oscillation parameters of the giants in the binary systems in order to determine their seismic fundamental parameters and find them to agree with the dynamic parameters from the literature if we adopt nonlinear scalings. We note that a curvature and glitch corrected $\Delta\nu_\mathrm{cor}$ should be preferred over a local or average value of $\Delta\nu$. We then compare the observed seismic parameters of the cluster giants to those scaled from independent measurements and find the same nonlinear behaviour as for the eclipsing binaries. Our final proposed scaling relations are based on both samples and cover a broad range of evolutionary stages from RGB to RC stars:  $g/\sqrt{T_\mathrm{eff}} = (\nu_\mathrm{max}/\nu_\mathrm{max,\odot})^{1.0075\pm0.0021}$ and $\sqrt{\bar\rho} = (\Delta\nu_\mathrm{cor}/\Delta\nu_\mathrm{cor,\odot})[\eta - (0.0085\pm0.0025) \log^2 (\Delta\nu_\mathrm{cor}/\Delta\nu_\mathrm{cor,\odot})]^{-1}$, where $g$, $T_\mathrm{eff}$, and $\bar\rho$ are in solar units, $\nu_\mathrm{max,\odot}=3140\pm5$\mh\ and $\Delta\nu_\mathrm{cor,\odot}=135.08\pm0.02$\mh , and $\eta$ is equal to one in case of RGB stars and $1.04\pm0.01$ for RC stars. }
{A direct consequence of these new scaling relations is that the average mass of stars on the ascending giant branch reduces to $1.10\pm0.03M$\sun\ in \na\ and $1.45\pm0.06M$\sun\ in \nb, allowing us to revise the clusters' distance modulus to $13.11\pm0.03$ and $11.91\pm0.03$\,mag, respectively. We also find strong evidence that both clusters are significantly older than concluded from previous seismic investigations.}

\authorrunning{Kallinger et al.}
\titlerunning{Nonlinear seismic scaling relations}
   \maketitle

\section{Introduction}	\label{sec:intro}
Mass and radius are amongst the most fundamental stellar parameters. Knowing them is essential to characterise a star, but also a prerequisite to characterise a planet orbiting that star. A frequently used method to determine stellar mass and radius is by comparing some observed properties, such as effective temperature and surface gravity, to those of stellar evolutionary tracks and derive a set of best-fit model parameters (mass, radius, age, etc.). This method is, however, quite uncertain not only because of the often large observational uncertainties but also because of large and frequently ignored systematic uncertainties in the models \citep[e.g.,][]{Stancliffe2016}. A direct measurement of stellar mass is possible through radial velocity measurements for stars with well known inclinations such as eclipsing, double-lined spectroscopic binaries (SB2), or through precise astrometry/interferometry of visually resolved wide binaries.  In such cases, the mass can be measured with high accuracy and precision (<3\%), but there are only a few such systems known.

Another method to infer stellar fundamental parameters is by means of asteroseismology \citep*[e.g., see the monograph of][]{aerts10}, in particular for cool stars with a convective surface layer (hence stars ranging from cool F to M on the main sequence and high up the giant branch). They are believed to show so-called solar-like oscillations, which are intrinsically damped and stochastically excited by the turbulence in the near-surface convection. Their oscillation spectrum presents a nearly regular frequency pattern of modes, which carry information about the internal structure of the star.

The simplest analysis of solar-like oscillations is based on the `global' properties of the oscillations, which are the large frequency separation, \dnu, between consecutive modes of the same spherical degree, $l$, and the frequency of the maximum oscillation power, \num, which we will call peak frequency for short. 

The large frequency separation is related to the inverse sound travel time across the stellar diameter, essentially scaling with the mean stellar density \den\ \citep[e.g.,][]{bro94} as,
\begin{equation} \label{eq:dnu}
\Delta\nu = \Big(2\int_0^R \frac{dr}{c}\Big)^{-1}\propto \sqrt{\bar\rho} \propto \sqrt{\frac{M}{R^3}},
\end{equation}
where $c$ is the sound speed and $M$ and $R$ are the stellar mass and radius, respectively. The peak frequency, however, is related to the acoustic cut-off frequency $\nu_{ac}$ in the stellar atmosphere \citep[e.g.,][]{bro91,kje95}, which in turn scales with the stellar surface gravity $g$ and effective temperature \teff\ as, 
\begin{equation} \label{eq:numax}
\nu_\mathrm{max} \propto \nu_{ac} \propto g / \sqrt{T_\mathrm{eff}} \propto \frac{M}{R^2 \sqrt{T_\mathrm{eff}}}.
\end{equation}
From this pair of scaling relations -- commonly called global asteroseismic scalings -- and a measurement for the effective temperature, one can estimate a star's surface gravity and mean density and subsequently its mass and radius by scaling \num\ and \dnu\ to those of the Sun \citep[e.g.,][]{kal10b,kal10a}. 

One might expect that the solar reference values are simple to measure but in practise a number of values can be found in the literature \citep[e.g.,][]{kje95,huber2011,mosser2013} that differ by more than the individual uncertainties. For the large separation this is largely because \dnu\ is in fact a function of frequency \citep[see, e.g.,][]{broomhall09} and therefore depends on the frequency range for which it is measured.  To measure the peak frequency requires to separate the oscillation signal from the underlying granulation background and unwarranted assumptions about the latter can bias the measurement \citep{kal14}. Using the seismic scaling relations therefore requires well-defined and consistently measured global oscillation parameters.

In practice, measurements of the asteroseismic global parameters have been extensively used to estimate masses and radii for stars with solar-like oscillations detected in CoRoT \citep{bag06} and \kep\ \citep{bor10} data. For more details we refer to the reviews by \cite{chaplin2013} and \cite{garcia2018}. Given the importance of the seismic scalings, several studies have tried to test their validity, but mostly focused on the $\Delta\nu - \bar\rho$ scaling because so far there is no secure theoretical basis for predicting \num\ \citep[e.g.,][]{belkacem2011}.  However, \cite{Viani2017} recently reported on a mean molecular weight and adiabatic index term in the $\nu_{ac}$ -- scaling of their stellar model.

The available studies that examine the \dnu --scaling follow one of two approaches: those testing the scalings with independently determined radii (for example from interferometry) and those confronting the scaling with model predictions. The observational approach shows generally good agreement \citep[within $\sim$4 and 16\% for main-sequence stars and red giants, respectively, see, e.g.,][]{huber2012,baines2014} but are limited by the uncertainties of the non-seismic benchmarks observations.  This is because, currently, seismic targets from CoRoT or \kep\ are typically quite faint. The theoretical approach \citep[e.g.,][]{white2011,guggenberger2016, sharma2016}, on the other hand, is in principle more accurate but has to deal with systematic effects in the models (such as the so-called ``surface effect''; e.g., \citealt{gruberbauer2013}, \citealt{Li2018}), which hamper a comparison between scaled \dnu\ and those determined from the model frequencies.

To validate seismic masses is even less practicable. In conclusion, even though the exquisite photometry of CoRoT and \kep\ allows to estimate seismic masses and radii to a precision of a few per cent, the underlying scaling relations can not be confirmed to better than about 4 and 16\% for main-sequence and red-giant stars, respectively. 

A promising way to improve the situation is given by eclipsing double-lined spectroscopic binary (eSB2) systems hosting at least one star with detectable solar-like oscillations. Complementary photometric observations of the eclipses and radial velocity measurements of the orbital motions allow to accurately constrain the orbital inclination and therefore the dynamic masses and radii of both components, even for quite faint stars. The first such reported system, KIC\,8410637, was detected by \cite{hekker2010}, and \cite{frandsen2013} found good agreement between the seismic and dynamic estimate of \lg\ for the oscillating red-giant component. However, \cite{huber2014} noted that the seismic and dynamic radius and mass deviate by about 9\% (2.7$\sigma$) and 17\% (1.9$\sigma$), respectively, indicating that the seismic scaling relations need some revision. 
\begin{figure}[b]
	\begin{center}
	\includegraphics[width=0.5\textwidth]{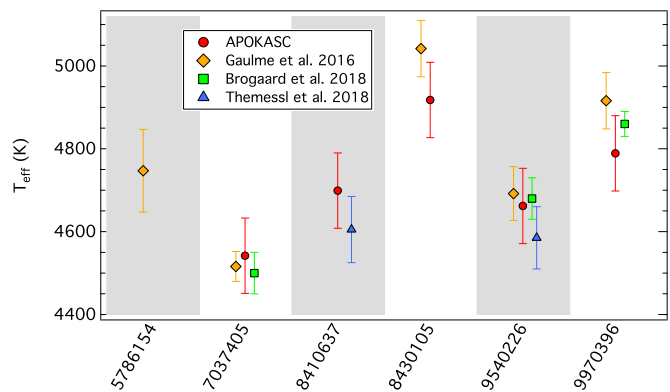}
	\caption{Comparison of \teff\ from various sources for our sample of red giants in eSB2 systems.} 
	\label{fig:temps} 
	\end{center} 
\end{figure}
This was confirmed by a recent study of \cite{gaulme2016}, who investigated a sample of 10 eSB2 systems observed by \kep\ with supplementary radial velocity observations, from which they could detect solar-like oscillations and measure dynamic masses and radii. They found that whereas the agreement between seismic and dynamic surface gravities is acceptable, the seismic scalings systematically overestimate the radii and masses by about 5 and 15\%, respectively, for various stages of red-giant evolution. An immediate consequence of this is that an age estimate based on the seismic mass \citep[via grid-modelling, e.g.,][]{Stello2009,Basu2010,kal10b,Quirion2010,Gai2011,Creevey2013,Hekker2013,Miglio2013,Serenelli2013} substantially underestimates the actual stellar age, which would be a serious issue for applications such as galactic archaeology. 

In this paper, we re-examine the seismic analysis of the key set of stars in the \cite{gaulme2016} sample and use the dynamic mass and radius measurements by \cite{gaulme2016}, \cite{Brogaard2017}, and \cite{themessl2016} to formulate new nonlinear\footnote{Strictly speaking, the original seismic scaling relations are power laws and are therefore already nonlinear. However, since there is no simple ``name'' for changing the power of $\bar\rho$, $g$, and $T_\mathrm{eff}$ from integer and half-integer numbers to rational numbers we prefer to call the original scalings \textit{linear} and the new scalings \textit{nonlinear}. 
} scaling relations for \num\ and \dnu. We then compare the observed seismic parameters of more than 60 red giants in the two open clusters \na\ and \nb\ to those scaled from independent measurements and find them to show the same nonlinear behaviour as the eclipsing binaries. We use this to refine the scaling coefficients and to extend the scalings to a broader range of evolutionary stages on the giant branch, including red clump stars. Finally, we discuss the resulting decrease of the average mass of stars on the red giant branch (RGB) of the two clusters and its consequences for the age determination.

\section{Gaulme's sample of oscillating red giants in eclipsing binaries} \label{sec:eb}

\citet[][hereafter G16]{gaulme2016} reported on a four-year radial velocity (RV) survey for a sample of eclipsing binary systems discovered by \kep. By combining the RV measurements and the photometric \kep\ observations of the eclipses, they determined the dynamical mass and radius for 14 double-lined spectroscopic binaries. Ten of these systems include a red giant that shows clear solar-like oscillations in the \kep\ photometry, for which they determined global seismic parameters. This allowed G16 to straight-forwardly test the seismic scaling relations with high precision and accuracy, for the first time.  We go beyond that and use their key set of stars to derive new seismic scalings for red giants. Since not all systems are suitable for such an analysis, we selected six out of the original ten binaries. The remaining four systems have either observations of insufficient accuracy or ambiguous evolutionary phases of the red giant components (see Sec.\,\ref{sec:seispar}). Our sample of stars is listed in Tab.\,\ref{tab:RMcalib}. 

\subsection{Defining our sample} \label{sec:definesample}
Two stars of our sample were also analysed by \citet[][hereafter T18]{themessl2016}, who reanalysed the spectroscopy obtained by \cite{frandsen2013} and \cite{Beck2014}. In the G16 analysis, the dynamic mass and radius of KIC\,8410637 are based on the original analysis of \cite{frandsen2013} using the eclipses covered by \kep\ between Q1 and Q10. With the full \textit{Kepler} data available, T18 refined the dynamic parameters. For KIC\,9540226, G16 had to fix the system inclination to 90\degr\ to help the fit to converge. T18 solved this problem and found a slightly lower inclination resulting in somewhat different but more reliable dynamic parameters. 

Two more stars were also investigated by \citet[][hereafter B18]{Brogaard2017}, who used additional spectroscopic observations to refine the dynamic solutions for KIC\,7037405 and KIC\,9970396. KIC\,9540226 was analysed by all three groups and while the mass estimates are in good agreement, the dynamic radius differs by about 1.6$\sigma$ between T18 and B18. We adopt the solution from T18 because they provide a more accurate mass but note that using the values from B18 does marginally affect our subsequent analysis. Summarising, for our sample of six stars, we adopt the dynamic parameters for two stars each from the original G16 analysis, from T18, and from B18, respectively. This diversity in source should make our analysis largely independent of any specific instrument (for the spectroscopy) and analysis method. All dynamic parameters are listed in Tab.\,\ref{tab:RMcalib}.

\subsection{Effective temperature scale}
A possible source for the discrepancy between the seismic and dynamic masses and radii is the temperature scale applied to the seismic scalings. \cite{Beck2018} pointed out that the stars, reported in the the sample of G16 are systematically hotter by a few 10K than stars in the sample of \cite{Beck2014}. They also concluded, that the difference is not big enough to reconcile the mass discrepancy between seismically and dynamically inferred masses. To cover the entire reported offsets one would, however, need to change \teff\ by about 10\% or 500\,K for a typical red giant with an effective temperature of 4800\,K, which seems not very plausible. In fact, the effective temperatures of our sample stars from various sources are generally in good agreement (see Fig.\,\ref{fig:temps}). We find a maximum difference of 127\,K for the reported \teff\ values, which can certainly not account for the discrepancy between the seismic and dynamic masses and radii. To be consistent with other \kep\ red giants we use the effective temperatures from the APOKASC Catalog \citep[DR13][]{pin2014} representing an asteroseismic and spectroscopic joint survey of stars in the \kep\ field. For KIC\,5786154 there is no \teff\ estimate available in the APOKASC Catalog, which is why we use the original value from G16 for this star. A detailed comparison with other \teff\ sources may be found in G16, showing a similar picture.

\begin{figure}[b]
	\begin{center}
	\includegraphics[width=0.5\textwidth]{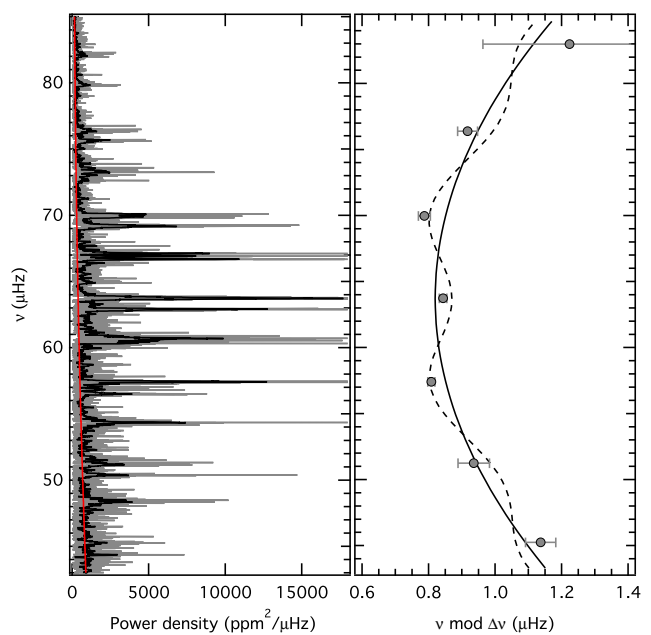}
	\caption{(Left) Original (grey line) and smoothed (black line) power density spectrum of KIC\,9970396. The red line indicates the granulation background. (Right) \'Echelle diagram of the significant radial modes. The solid and dashed lines correspond to our asymptotic fit (Eq.\,\ref{eq:glitch}) and the curvature term of the fit, respectively.} 
	\label{fig:KIC4663623} 
	\end{center} 
\end{figure}

\begin{figure}[t]
	\begin{center}
	\includegraphics[width=0.5\textwidth]{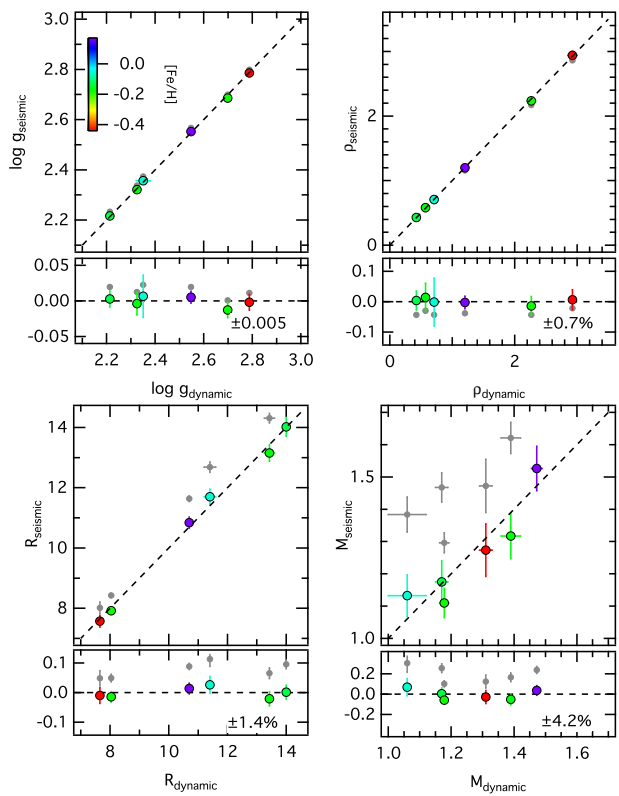}
	\caption{Seismic versus dynamic parameters for our sample of seven eSB2 binaries. The four main panels show the direct comparison for each of the four parameters $\log g$, $\bar\rho$, radius, and mass, with the dashed lines indicating the 1:1 relation. $g$ is in cgs units, $\bar\rho$ is in units of $10^{-3}\bar\rho\sun$, and $R$ and $M$ are in solar units. The coloured symbols correspond to seismic values based on the new nonlinear scalings (Eq.\,\ref{eq:numDynRelation} and \ref{eq:nonlinear_dnu}) with \num\ and $\Delta\nu_\mathrm{cor}$ as seismic input. The colour indicates the metallicity [Fe/H]. The grey symbols correspond to the results of classical seismic scaling using the Sun as reference and \num\ and $\Delta\nu_\mathrm{c}$ as seismic input. The small panels under each main panel show the relative difference between the seismic and dynamic parameters in the sense of $(seismic-dynamic)/dynamic$. The numbers indicate the mean absolute deviation from perfect agreement for the new scalings (see Tab.\,\ref{tab:RMcalib}).} 
	\label{fig:RMcalib} 
	\end{center} 
\end{figure}

\section[Re-examing the global seismic parameters]{Re-examing the global seismic parameters and their associated fundamental parameters} \label{sec:seispar}

All photometric time series used in this analysis have been obtained with the NASA \kep\ space telescope spanning from quarter 2 to 17 with a total length of about 1420 days. The \kep\ raw SAP-FLUX measurements\footnote{available at the MAST website: \href{url}{https://archive.stsci.edu/kepler/}} were smoothed with a triangular filter to suppress intrinsic and instrumental long-term trends with time scales longer than about 10d. Even though \kep\ is supposed to observe continuously, there are several instances where scientific data acquisition is affected, producing quasi-regular gaps \citep{garcia2014}. Gaps that are shorter than about 3/\num\ are filled by linear interpolation improving the duty cycle from about 93 to about 95\%. This is sufficient to correct for most of the objectionable regularities in the spectral window function \citep{kal14}. 

In a first step, we re-examined the seismic analysis for the G16 sample following the methods described in \cite{kal10b,kal12,kal14} and find sufficiently accurate results for seven of the original ten stars. For KIC\,7377422 the oscillation modes are too weak to make any firm conclusions. For KIC\,10001167 and KIC\,4663623 the seismic and dynamic uncertainties, respectively, are too large to include the stars in the further analysis. We do, however, note that excluding the latter two stars does only marginally change the main result of our analysis but improves the final uncertainties.  

\subsection{The peak frequency \num } \label{sec:numax}

To determine the peak frequency we adopt the approach of \cite{kal14} and fit a global model to the observed power density spectra, where we restrict the spectra to between 0.1\num\ and the Nyquist frequency. The model consists of a flat noise component, two super-Lorentzian functions with an exponent fixed to four, and a Gaussian with the central frequency representing \num . For the fit we use the Bayesian nested sampling algorithm \textsc{MultiNest} \citep{feroz2009,feroz2013}, which delivers posterior probability distributions for the parameters estimation and a so-called global model evidence. The latter is a statistical measure that allows us to reliably rate if a given model represents the observations better than some other model. 

The resulting peak frequencies agree with those reported by G16 on average to within 2$\sigma$. For five of the six stars, however, our \num\ is larger than those from G16 (by on average 1.6\%). The reason for this is likely due to a different treatment of the background signal. \cite{kal14} systematically compared various background models and found systematic differences in \num\ depending on the specific treatment of the granulation signal. Even though G16 used the same model for the granulation signal as we do, they added a coloured-noise component \citep[as suggested by][]{kal14}. To further investigate this we repeated the fits now following the exact approach of G16 (Gaulme, private communication) and compared the resulting model evidences delivered by \textsc{MultiNest}. Interestingly, the additional coloured-noise component is statistically justified only for KIC\,9540226. For the other five stars it represents an overfitting of the data presumably distorting the granulation model and causing a small but systematic underestimation of \num . We therefore prefer our values of \num\ for the subsequent analysis (see Tab.\,\ref{tab:RMcalib}).

\subsection{The central large separation \dc} \label{sec:dnu_central}

For the large frequency spacing G16 follow the approach of \cite{mosser2011} to compute an average \dnu\ from the universal pattern, which is then converted into an asymptotic \dnu\ using the semi-empirical relation of \cite{mosser2013} (see Sec.\,\ref{sec:asymDnu}). They thereby use a fixed conversion factor (1+$\zeta$), with $\zeta=0.038$. Such a fixed conversion does, however, not account for the observed variety of the mode curvature of up to some ten per cent \citep[see Fig.\,6 in][]{mosser2013}, which add an additional uncertainty of 1-2\% to the random error of typically 0.1\% in \dnu\ as given by G16.

\begin{table*}[t]
\begin{small}
\centering
\caption{Stellar parameters for the six known eSB2 in the \kep\ sample from dynamical modeling and asteroseismic scaling. The parameters $M$ (mass), $R$ (radius), and $\bar\rho$ (mean density) are in solar units ($10^{-3}\bar\rho\sun$ in case of the $\bar\rho$). The effective temperatures, \teff, are taken from the APOKASC Catalogue, except for $^{(d)}$, which is taken from \cite{gaulme2016}. The peak frequency, \num, and the large frequency separations \dc\ (central), \dcor\ (corrected from Eq.\,\ref{eq:glitch}), and $\Delta\nu_\mathrm{as}$ (asymptotic), are in \mh. The last two columns give the statists with $\bar\sigma$ corresponding to the average uncertainty of the parameter and MAD being the mean absolute deviation of the relative difference between the dynamic and asteroseismic parameter.
\label{tab:RMcalib}}
\begin{tabular}{l|   r  r  r  r  r   r  |r  r}
\hline\hline
\noalign{\smallskip}
KIC						&			5786154\tablefootmark{a}	&	7037405\tablefootmark{b}			&	8410637\tablefootmark{c}			&	8430105\tablefootmark{a}	&	9540226\tablefootmark{c}			&	9970396\tablefootmark{b}	&$\bar\sigma$& MAD\\
\noalign{\smallskip}
\hline
\noalign{\smallskip}
$T_\mathrm{eff}$ 	&	4747$\pm$100	\tablefootmark{d}        		&	4542$\pm$91		&	4699$\pm$91		&	4918$\pm$91			&	4662$\pm$91		&	4789$\pm$91&92\,K&\\
\num			   	        &	29.84$\pm$0.18	&	22.12$\pm$0.13	&	46.87$\pm$0.23	&	78.14$\pm$0.91		&	27.73$\pm$0.16	&	63.00$\pm$0.26&0.7\%&\\
\dnu$_\mathrm{c}$		&	3.513$\pm$0.009	&	2.784$\pm$0.010	&	4.648$\pm$0.011	&	7.143$\pm$0.029		&	3.196$\pm$0.012	&	6.302$\pm$0.010&0.3\%&\\
\dnu$_\mathrm{cor}$		&      3.520$\pm$0.011	&	2.726$\pm$0.011 	&	4.594$\pm$0.005	&	7.225$\pm$0.071		&	3.181$\pm$0.011	&	6.289$\pm$0.009&0.3\%&\\
\dnu$_\mathrm{as}$		&      3.762$\pm$0.041	&	3.000$\pm$0.047 	&	4.960$\pm$0.025	&	7.908$\pm$0.187		&	3.413$\pm$0.042	&	6.504$\pm$0.069&1.3\%&\\
\noalign{\smallskip}
\hline
\noalign{\smallskip}
				\multicolumn{9}{c}{dynamical modeling}\\
\noalign{\smallskip}
\hline
\noalign{\smallskip}
log\,$g$		&	2.350$\pm$0.029	&	2.214$\pm$0.009	&	2.547$\pm$0.006	&	2.788$\pm$0.009		&	2.325$\pm$0.015	&	2.699$\pm$0.010	&0.012&\\
$\bar\rho$	        &	0.71$\pm$0.05		&	0.43$\pm$0.01		&	1.20$\pm$0.02		&	2.92$\pm$0.07			&	0.57$\pm$0.03		&	2.27$\pm$0.07		&3.6\%&\\
M			&	1.06 $\pm$0.06	        &	1.17 $\pm$0.02	        &	1.47 $\pm$0.02	        &	1.31 $\pm$0.02	          	&	1.39 $\pm$0.03	         &	1.18 $\pm$0.02 	&2.2\%&\\
R			&	11.40$\pm$0.20	&	14.00$\pm$0.09	&	10.60$\pm$0.05	&	 7.65$\pm$0.05	                 &	13.4$\pm$0.2	         &	 8.04$\pm$0.07	        &0.9\%&\\
\noalign{\smallskip}
\hline
\noalign{\smallskip}
				\multicolumn{9}{c}{new nonlinear asteroseismic scaling (input: $T_\mathrm{eff}$, \num , and \dnu$_\mathrm{cor}$)}\\
\noalign{\smallskip}
\hline
\noalign{\smallskip}
log\,$g$		&	2.357$\pm$0.007	&	2.216$\pm$0.006	&	2.552$\pm$0.006	&	2.786$\pm$0.007		&	2.321$\pm$0.006	&	2.686$\pm$0.005	&0.006&0.005\\
$\bar\rho$   	&	0.71$\pm$0.01		&	0.43$\pm$0.01	       &	1.20$\pm$0.01		&	2.94$\pm$0.06			&	0.58$\pm$0.01		&	2.23$\pm$0.02		&1.5\%&0.7\%\\
M			&	1.13$\pm$0.06		&	1.18$\pm$0.07	       &	1.53$\pm$0.07		&	1.32$\pm$0.07			&	1.32$\pm$0.07		&	1.11$\pm$0.05		&5.2\%&4.2\%\\
R  			&	11.70$\pm$0.26	&	14.02$\pm$0.31	&	10.84$\pm$0.18	&	7.57$\pm$ 0.20   		&	13.15$\pm$0.27	&	7.92$\pm$0.12		&2.1\%&1.4\%\\	 
\noalign{\smallskip}
\hline
\end{tabular}
\tablefoot{
Dynamic parameters are taken from:
\tablefoottext{a}{\cite{gaulme2016}}--
\tablefoottext{b}{\cite{Brogaard2017}}--
\tablefoottext{c}{\cite{themessl2016}}\\
}
\end{small}
\end{table*}

We think that the use of a scaled asymptotic \dnu\ in a star-by-star analysis, rather than in an ensemble sense, does in fact impair rather than improve the subsequent analysis if the departure from regular frequency spacing in the form of `curvature' and `glitch' properties are not measured for each star (see Sec.\,\ref{sec:asymDnu}).  This is why we prefer to use an observed value of \dnu. More specifically, we start our analysis with the average frequency spacing of the three central radial orders \dnu$_c$, which is determined by fitting a sequence of eight Lorentzian profiles to the power density spectra. The model covers three consecutive orders of $l$ = 0 and 2 modes and the intermediate $l$ = 1 modes, whose frequencies are parameterised by the frequency of the central radial mode and a large and small frequency separation. For the fit we use again \textsc{MultiNest} and refer to \cite{kal10b,kal12} for more details about the automated approach. The seismic parameters and their uncertainties are listed in Tab.\,\ref{tab:RMcalib}.

\subsection{Evolutionary stage}
An often used method to evaluate the evolutionary state of a red giant is by means of the period spacing of its mixed dipole modes \citep[e.g.,][]{bedding2011,Mosser2014}. For many of the stars in our sample this is, however, not possible because either the frequency resolution or the signal-to-noise ratio of the modes is not sufficient. We therefore determine the evolutionary stage from the phase shift of the central radial mode \citep{kal12,ChristensenDalsgaard2014} and find that all but one star in the original G16 sample are ascending RGB stars. Only for KIC\,9246715 we cannot clearly determine if it is a H-shell burning star or a core-He burning star that has not experienced a helium flash on the tip of the giant branch (also called a secondary clump star). \cite{gaulme2014} and later on \cite{Rawls2016} classified this star as a secondary clump star, which is why we remove it from our sample. We further note that we find an independent classification only for KIC\,9970396, which is identified as RGB star by \cite{Elsworth2017}. 

\subsection{The asymptotic large separation $\Delta\nu_\mathrm{as}$} \label{sec:asymDnu}

The frequencies of low-degree and high-order acoustic modes approximately follow an asymptotic relation \citep{tassoul1980,gough1986}, which reduces in its simplest form for radial modes ($l$ = 0) to, 
\begin{equation} \label{eq:asymrel1}
\nu_{(n,0)} \simeq \nu_\mathrm{c} + \Delta\nu\,(n-n_\mathrm{c}),
\end{equation}
when linked to the frequency of the central radial mode of the oscillation power excess (meaning the radial mode closest to \num), where $n$ is a mode's radial order and the subscript $c$ denotes values for the central radial mode. This form can clearly not account for an accurate description of the observed radial mode pattern in red giants. They show a noticeable curvature in the \'echelle diagram (Fig.\,\ref{fig:KIC4663623}) at all evolutionary stages \citep[e.g.,][]{kal12} so that higher order terms in the asymptotic expansion need to be considered. 

Following \cite{mosser2011}, the second-order development of the asymptotic theory can be described with a curvature term $\alpha$:
\begin{equation} \label{eq:asymrel0}
\nu_{(n,0)} \simeq \nu_\mathrm{c} + \Delta\nu_\mathrm{cor} \, \Big[n-n_\mathrm{c} + \frac{\alpha}{2}(n-n_\mathrm{c})^2\Big],
\end{equation}
where \dcor\ represents the corrected frequency difference between radial modes at the central radial mode. Strictly speaking, Eq.\,\ref{eq:dnu} is, however, defined for the large separation at infinitely high frequencies and not at \num, so that \dcor\ needs to be linked to the asymptotic large separation $\Delta\nu_\mathrm{as}$ \citep{mosser2013} as: 
\begin{equation} \label{eq:dnu_as}
\Delta\nu_\mathrm{as} =  \Delta\nu_\mathrm{cor}\, \Big[1 + \frac{\alpha \,n_\mathrm{c}}{2} \Big].
\end{equation}
The curvature term is clearly responsible for the difference between the observed and asymptotic values of the large separation and correcting for it should improve the $\Delta\nu - \bar\rho$ scaling. 

\cite{mosser2013} measured the curvature terms for a few red giants to determine a relation $\alpha$(\dnu ), which intends to convert an observed value of \dnu\ into a physically-grounded asymptotic value without the need for an in-depth analysis of the mode pattern. The scaling is, however, only based on a small number of stars and does not account for the relatively large scatter of $\alpha$ independent of \dnu. Hence, using  $\alpha$(\dnu ) adds significant uncertainties to $\Delta\nu_\mathrm{as}$. This might be acceptable in a statistical sense for a large sample of stars but certainly not for the detailed analysis of our sample of eSB2 stars. We therefore intend to measure the curvature term individually. To do so, another effect needs to be taken into account.

\cite{vrard2015} have shown that the observed deviations from Eq.\,\ref{eq:asymrel0} can be characterised in terms of the location of the second ionization zone of helium \citep[e.g.,][]{houdek2007}. They use the second-order expansion to correct the observed frequencies for curvature and fit a harmonic term to the residuals. Here we are more interested in the curvature term and therefore expand the asymptotic relation to: 
\begin{equation} \label{eq:glitch}
\nu_{(n)} = \nu_\mathrm{c} + \Delta\nu_\mathrm{cor} \, \Big[n - n_\mathrm{c} + \frac{\alpha}{2}(n-n_\mathrm{c})^2 + A\sin\Big(2\pi\,\Big(\frac{n-n_\mathrm{c}}{G}+\phi_\mathrm{c}\Big)\Big) \Big],
\end{equation}
where the harmonic term accounts for the periodic frequency modulation caused by the second helium ionization zone, often called the glitch signal. $A$, $G$, and $\phi_\mathrm{c}$ are the glitch term's amplitude, period, and phase, respectively, at the central radial mode. We note here that it is not necessary to specifically take care of the crosstalk between parameters, like $\alpha$ and $G$. Any possible crosstalk will be reflected in the posterior probability distributions delivered by \textsc{MultiNest}, from which we determine the most likely parameters and their uncertainties. We further note that $\alpha$ from Eq.\,\ref{eq:asymrel0} and Eq.\,\ref{eq:glitch} turn out to be practically identical.

To determine the curvature and the glitch parameters we first need the frequencies of the radial modes.
Our method to determine \dc\ (Sec.\,\ref{sec:dnu_central}) also provides an estimate of $\nu_\mathrm{c}$ \citep[see][for more details]{kal10b,kal12}, which we use to predict the initial positions of the radial modes. To extract their frequencies, we fit Lorentzian profiles to the power density spectrum again using \textsc{MultiNest} and consider a mode as statistically significant if the model evidence provided by \textsc{MultiNest} is at least ten times larger than for a fit with a straight line (meaning only noise)\footnote{In probabilty theory an odds ratio of 10:1 is considered as strong evidence \citep{jeffreys98}.}. We then fit Eq.\,\ref{eq:glitch} to the extracted mode frequencies to determine the curvature and glitch parameters. We again fit various models (straight line, Eq.\,\ref{eq:asymrel0}, and \ref{eq:glitch}) to determine the significance of the result. The resulting \dcor\ and $\Delta\nu_\mathrm{as}$ values are listed in Tab.\,\ref{tab:RMcalib}. As an example, Fig.\,\ref{fig:KIC4663623} shows the power density spectrum and \'echelle diagram of the radial modes for KIC\,9970396 along with the best fit. More details about the mode fitting and the curvature and glitch analysis are subject to a separate paper (Kallinger et al., in preparation).  

\subsection{Solar-calibrated seismic fundamental parameters from \num\ and \dc}

By comparing \dnu\ and \num\ to the respective solar reference values we can now easily determine the mean stellar density and surface gravity, and subsequently the mass and radius, from Eq.\,\ref{eq:dnu} and \ref{eq:numax}. To ensure a consistent internal calibration, the solar low-degree oscillation spectrum needs to be analysed with the same tools as used for the stellar power density spectra. We refer to \dnu$_\mathrm{c,\odot} = 134.88\pm0.04$\mh\ \citep{kal10b} and \num$_{,\odot} = 3140\pm4$\mh\ \citep{kal14} as solar reference values, which are both determined from one-year time series from the green channel of the SOHO/VIRGO data \citep{Frohlich1997} during solar minimum 23. The resulting seismic masses and radii of our set of stars have internal uncertainties of about 3.8 and 2.0\%, respectively. A comparison between the dynamic and solar-calibrated seismic parameters is shown in Fig.\,\ref{fig:RMcalib} (grey symbols). We find rather small deviations between the seismic and dynamic surface gravities (0.014$\pm$0.008) and mean densities ($-2.5\pm$1.8\%). They do, however, translate into significant deviations when converted into $R$ (6.4$\pm$2.6\%) and $M$ (17$\pm$7\%), which is fully consistent with the result of G16 (see Tab.\,\ref{tab:RMsystematic}).

\subsection{Model-calibrated seismic fundamental parameters from \num\ and \dc}

\cite{white2011} and only recently \cite{guggenberger2016} and \cite{sharma2016} suggested to correct the observed \dnu\ with an empirical function, which they deduce from stellar models. They found that the computed radial eigenfrequencies of stellar models have a large separation that does not exactly follow the $\Delta\nu-\bar\rho$ scaling. The deviations are on the order of up to a few percent and depend on the effective temperature and chemical composition of the models. We apply their corrections to \dnu$_\mathrm{c}$ and compute the resulting mean density (Tab.\,\ref{tab:RMsystematic}), which in fact sufficiently aligns the $\Delta\nu-\bar\rho$ scaling.  However, if we compute the stellar radius and mass from this, we still find systematic deviations of 2 -- 4\% and 8 -- 12\%, respectively. This is because also the $\nu_\mathrm{max}-g$ scaling shows a systematic deviation. 

Now, an obvious problem in such model-dependent approaches is that the model frequencies suffer from systematic errors, such as the so-called surface effect \citep[e.g.,][]{kjeldsen2008}, which is an offset between observed and computed oscillation frequencies. For the Sun, the offset is known to increase with frequency and hence affects the large separation with \dnu\ being about 1\% larger in the models than observed \citep[e.g.,][]{Houdek2017}. For more evolved star, \cite{Sonoi2015}, \cite{Ball2017}, and only recently \cite{Li2018} suggested surface-effect corrections, from which we conclude that the model \dnu\ is affected on the order of 1--2\%. Furthermore, \cite{gruberbauer2013} found evidence that the surface effect significantly varies across the HR-diagram and can even be `anti-solar'. We therefore think that the \dnu\ corrections that arise from stellar models are rather connected to the surface effect in stellar models but do not in their current form reliably reflect the $\Delta\nu-\bar\rho$ scaling of real stars. 
%

\begin{table}[t]
\begin{tiny}
\centering
\caption{Average systemtatic deviation, $\chi$, and its rms scatter, between dynamic and seismic parameters for different input \dnu\ and reference values. The first column gives the indicies for the applied reference values. The actual values are listed below. 
\label{tab:RMsystematic}}
\begin{tabular}{llrrrr}
\hline\hline
\noalign{\smallskip}
reference&input&$\chi_{\log g}$&$\chi_{\bar\rho}$&$\chi_{R}$&$\chi_{M}$\\
(\num\ | \dnu\ )&$\Delta\nu$&	&[\%]&[\%]&[\%]\\
\noalign{\smallskip}
\hline
\noalign{\smallskip}
				\multicolumn{6}{c}{solar-calibrated}\\
\noalign{\smallskip}
\hline
\noalign{\smallskip}
($_{\odot}$ | $_\mathrm{c,\odot}$)&\dnu$_\mathrm{c}$	& 0.014$\pm$0.008 &	-2.5$\pm$1.8	&	6.4$\pm$2.6		&	17$\pm$7\\
($_{\odot}$ | $_\mathrm{cor,\odot}$)&\dnu$_\mathrm{cor}$	& 0.014$\pm$0.008 &	-3.7$\pm$0.9	&	7.6$\pm$2.7		&	20$\pm$8\\
($_{\odot}$ | $_\mathrm{as,\odot}$)&\dnu$_\mathrm{as}$	& 0.014$\pm$0.008 &	6.7$\pm$5.0	&	-3.0$\pm$4.3		&	-3$\pm$9\\
\noalign{\smallskip}
\hline
\noalign{\smallskip}
				\multicolumn{6}{c}{$\Delta\nu$ calibrated according to \cite{guggenberger2016}}\\
\noalign{\smallskip}
\hline
\noalign{\smallskip}
($_{\odot}$ | )&\dnu$_\mathrm{c}$	                  & 0.014$\pm$0.008 &	-0.5$\pm$2.8	&	4.2$\pm$3.1		&	12$\pm$8\\
($_\mathrm{RGB}$ | )&\dnu$_\mathrm{c}$	& 0.000$\pm$0.008 &	-0.5$\pm$2.8	&	0.8$\pm$2.9		&	1$\pm$7\\
\noalign{\smallskip}
\hline
\noalign{\smallskip}
				\multicolumn{6}{c}{$\Delta\nu$ calibrated according to \cite{white2011}}\\
\noalign{\smallskip}
\hline
\noalign{\smallskip}
($_{\odot}$ | )&\dnu$_\mathrm{c}$	& 0.014$\pm$0.008 &	0.2$\pm$3.7	&	3.6$\pm$3.8		&	11$\pm$9\\
($_\mathrm{RGB}$ | )&\dnu$_\mathrm{c}$	& 0.000$\pm$0.008 &	0.2$\pm$3.7	&	0.2$\pm$3.7		&	0$\pm$8\\
\noalign{\smallskip}
\hline
\noalign{\smallskip}
				\multicolumn{6}{c}{$\Delta\nu$ calibrated according to \cite{sharma2016}}\\
\noalign{\smallskip}
\hline
\noalign{\smallskip}
($_{\odot}$ | )&\dnu$_\mathrm{c}$	                  & 0.014$\pm$0.008 &	-1.3$\pm$2.9 	&	2.0$\pm$3.0		&	8$\pm$7\\
($_\mathrm{RGB}$ | )&\dnu$_\mathrm{c}$	& 0.000$\pm$0.008 &	-1.3$\pm$2.9	&	-1.2$\pm$2.8		&	-2$\pm$7\\
\noalign{\smallskip}
\hline
\noalign{\smallskip}
				\multicolumn{6}{c}{RGB-calibrated}\\
\noalign{\smallskip}
\hline
\noalign{\smallskip}
($_\mathrm{RGB}$ | $_\mathrm{c,RGB}$)&\dnu$_\mathrm{c}$	& 0.000$\pm$0.008 &	0.1$\pm$1.8	&	0.1$\pm$2.5		&	-0.1$\pm$6.3\\
($_\mathrm{RGB}$ | $_\mathrm{cor,RGB}$)&\dnu$_\mathrm{cor}$	& 0.000$\pm$0.008 &	0.0$\pm$1.0	&	0.1$\pm$2.4		&	-0.2$\pm$6.6\\
($_\mathrm{RGB}$ | $_\mathrm{as,RGB}$)&\dnu$_\mathrm{as}$	& 0.000$\pm$0.008 &	0.0$\pm$4.8	&	0.0$\pm$4.5		&	-0.5$\pm$8.7\\
\noalign{\smallskip}
\hline
\noalign{\smallskip}
				\multicolumn{6}{c}{$g/\sqrt{T_\mathrm{eff}}=(\nu_\mathrm{max}/\nu_\mathrm{max,\odot})^\kappa$ and $\sqrt{\rho} = \Delta\nu /  \Delta\nu_{\odot} [1- \gamma \log^2(\Delta\nu/\Delta\nu_{\odot})]^{-1}$}\\
\noalign{\smallskip}
\hline
\noalign{\smallskip}
($_\mathrm{\odot}^{\kappa}$ | $_\mathrm{c,ref}$)$^{*}$&\dnu$_\mathrm{c}$	& 0.000$\pm$0.007 &	-0.6$\pm$2.1	&	0.7$\pm$2.5		&	1.1$\pm$4.5\\
($_\mathrm{\odot}^{\kappa}$ | $_\mathrm{cor,ref}$)$^{**}$&\dnu$_\mathrm{cor}$	& \textbf{0.000$\pm$0.007} &	\textbf{0.1$\pm$1.0}	&	\textbf{0.0$\pm$1.7}		&	\textbf{-0.5$\pm$3.4}\\
\noalign{\smallskip}
\hline
\noalign{\smallskip}
				\multicolumn{3}{l}{\,\,\,\,$\kappa = 1.0080\pm0.0024$}		&\multicolumn{3}{r}{$\nu_\mathrm{max,\odot}$ = 3140$\pm$5\mh}\\
				\multicolumn{3}{l}{$^{*}$ $\gamma = 0.0043\pm0.0025$}		&\multicolumn{3}{r}{$\nu_\mathrm{max,RGB}$ = 3245$\pm$50\mh}\\
				\multicolumn{3}{l}{$^{**}$$\gamma = 0.0085\pm0.0025$}	&\multicolumn{3}{r}{$\Delta\nu_\mathrm{c,\odot}$ = 134.89$\pm$0.04\mh}\\
				\multicolumn{6}{r}{$\Delta\nu_\mathrm{cor,\odot}$ = 135.08$\pm$0.02\mh}\\
				\multicolumn{6}{r}{$\Delta\nu_\mathrm{as,\odot}$ = 137.9$\pm$0.3\mh}\\
				\multicolumn{6}{r}{$\Delta\nu_\mathrm{c,RGB}$ = 133.1$\pm$1.3\mh}\\
				\multicolumn{6}{r}{$\Delta\nu_\mathrm{cor,RGB}$ = 132.5$\pm$1.3\mh}\\
				\multicolumn{6}{r}{$\Delta\nu_\mathrm{as,RGB}$ = 142.2$\pm$1.6\mh}\\
\end{tabular}
\end{tiny}
\end{table}

\subsection{Solar-calibrated seismic fundamental parameters from \num\ and \dnu$_\mathrm{as}$}

To determine a new set of seismic fundamental parameter, which are now based on \num\ and \dnu$_\mathrm{as}$, first requires an asymptotic value for the solar large separation. Applying the same procedure as in Sec.\,\ref{sec:asymDnu} to solar data gives $\Delta\nu_\mathrm{as,}\sun=137.94\pm0.32$\mh , which agrees well with the value derived by \cite{mosser2013} -- see their Tab.1. 

The seismic surface gravities stay unchanged but while the \dnu$_\mathrm{c}$-based mean stellar densities underestimate the dynamic values on average by about 2.5\%, the new \dnu$_\mathrm{as}$-based densities overestimate the dynamic ones by almost 7\% (see Tab.\,\ref{tab:RMsystematic}). Since $M\propto g^2/\bar\rho^2$ and $R \propto g/\bar\rho^2$, the resulting seismic masses and radii from \dnu$_\mathrm{as}$ are smaller than the \dc $-$based values and do in fact agree well with the dynamic estimates with only small systematic differences (see Tab.\,\ref{tab:RMsystematic}).

\cite{mosser2013} argued that the asymptotic large separation is a physically more meaningful representation of the mean stellar density than an observed value of \dnu\ and should therefore result in more realistic masses and radii \citep[see also][]{Belkacem2013}. We agree on the first part of the argument and clearly the \dnu$_\mathrm{as}-$based masses and radii agree better with the dynamic measurements than the \dnu$_\mathrm{c}-$based results do (see Tab.\,\ref{tab:RMsystematic}). However, this apparent agreement is because the systematic deviations in $g$ and $\bar\rho$, which represent the misaligned scaling between the seismic measurements and the physical properties of the star, almost cancel out when $M$ and $R$ are computed. The improvements in $M$ and $R$ are therefore largely due to the mathematical characteristic of the scalings. In fact the $\Delta\nu_\mathrm{as}-\bar\rho$ scaling is even more misaligned than the $\Delta\nu_\mathrm{c}-\bar\rho$ but in the opposite direction. We think the reason for this is due to the solar reference values and that the scalings are not strictly homologous.

\section{The nonlinearity of the scaling relations} \label{sec:newrefs}

Using the Sun as a reference for red giants implicitly assumes that the interior structure of a red giant scales linearly with the solar structure; meaning that the two are homologous. This is clearly not the case and it is therefore not surprising to find inherent systematic inaccuracies in the seismic-based fundamental parameters of red giants, such as mass and radius. A better approach would be to scale the seismic observables relative to those of one or more stars of similar interior structure. Our sample of eSB2 stars allows this for the first time with sufficient accuracy.

\begin{figure}
	\begin{center}
	\includegraphics[width=0.5\textwidth]{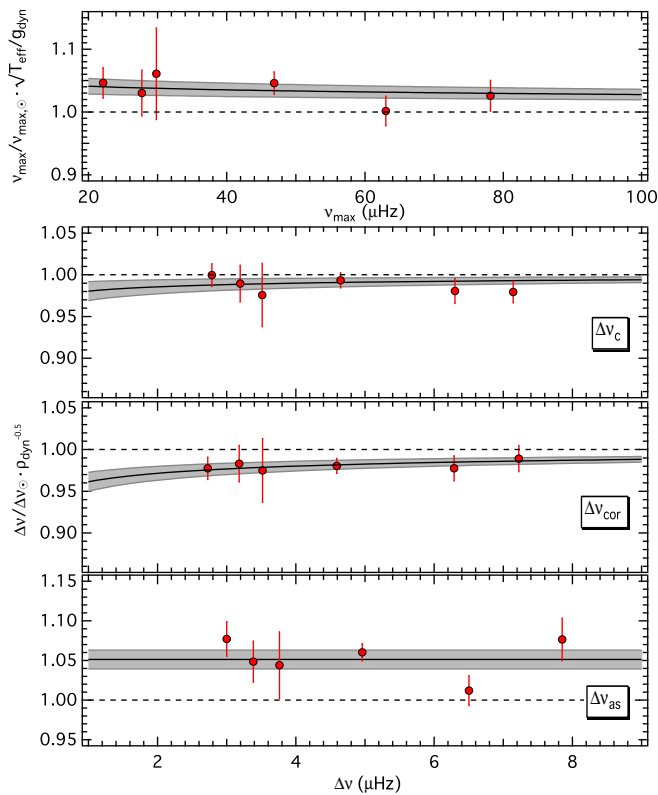}
	\caption{Reference values for \num\ (top panel) and \dnu\ (bottom three panels) that result from the ratio of the seismic and dynamic parameters (in solar units). The dashed lines indicate the solar reference value. The full black line lines show the new reference values and the grey-shaded range their uncertianties. 
}
	\label{fig:RefValues} 
	\end{center} 
\end{figure}

For the peak frequency we follow Eq.\,\ref{eq:numax} and compare the measured \num\ to $g_\mathrm{dyn}/\sqrt{T_\mathrm{eff}}$ (all in solar units) and find the resulting ratio to differ significantly from the solar reference, with an average value of \num$_\mathrm{,RGB} = 3245\pm50$\mh\ (see Fig.\,\ref{fig:RefValues}, top panel). This indicates that the solar peak frequency is not well-suited as a reference for red giants. Strictly speaking, this might only be valid for RGB stars with a \num\ in the range covered by the eSB2 stars (about 20 to 80\mh).
For a more general approach we fit the dynamic parameters with, 
\begin{equation}
\label{eq:numDynRelation}
\frac{g_\mathrm{dyn}} {\sqrt{T_\mathrm{eff}}} = \Big (\frac{\nu_\mathrm{max}}{\nu_\mathrm{max,\odot}} \Big ) ^{\kappa}.
\end{equation}
For the fit we use again \textsc{MultiNest}, which like any other fitting algorithm can only consider uncertainties in the dependent variable; in our case $g_\mathrm{dyn}/\sqrt{T_\mathrm{eff}}$). To also account for the uncertainties of the independent variable (\num$/\nu_\mathrm{max,\odot}$) we first add normal distributed uncertainties to the dependent variable with $\sigma$ being equal to the actual uncertainties of the individual measurements, and then redo the fit, adding up a probability density distribution for the exponent, $\kappa$. We iterate this procedure until the fitting parameter converges to a value that stays within one per cent of the corresponding uncertainty. The resulting accumulated probability density distribution is then fit by a Gaussian in order to determine the most likely parameter and its uncertainty. We find $\kappa = 1.0080 \pm 0.0024$, which is more than 3$\sigma$ different from unity, representing linear scaling. Even though this is only based on six stars we note that the Bayesian model evidence for the nonlinear \num-scaling is significantly better than for the linear scaling with a fixed \num$_\mathrm{,RGB}$. This makes also more sense from a physical point of view as Eq.\,\ref{eq:numDynRelation} does smoothly connect the giant branch to the solar regime, where $(\nu_\mathrm{max}/\nu_\mathrm{max,\odot})$ approaches one.

For the large separation we find a similar situation (see Fig.\,\ref{fig:RefValues} for \dc\ and \dcor ). As for the peak frequency, an average reference value \dnu$_\mathrm{c,RGB}$ = 133.1$\pm$1.3\mh , determined from our six eSB2 stars, improves the alignment of the linear $\Delta\nu - \bar\rho$ scaling but is again statistically insignificant compared to a power law of the form $\sqrt{\bar\rho} \propto \Delta\nu^{\,\gamma} $. However, we find that the function,
\begin{equation}	\label{eq:nonlinear_dnu}
\Delta\nu = \Delta\nu_\mathrm{ref} \cdot \sqrt{ \bar\rho_\mathrm{dyn}} = \Delta\nu_{\odot}\, \left [1- \gamma \log^2{(\Delta\nu/\Delta\nu_{\odot})} \right ] \cdot \sqrt{ \bar\rho_\mathrm{dyn}}
\end{equation}
gives an even better statistical evidence, with an odds ratio of about 14:1. The \textsc{MultiNest} fit  gives $\gamma$ = 0.0043$\pm$0.0025 and 0.0085$\pm$0.0025 for \dnu$_\mathrm{c}$ and \dnu$_\mathrm{cor}$, respectively, again considering the uncertainties of all parameters. Like for \num\ (Eq.\,\ref{eq:numDynRelation}), the newly established $\Delta\nu_\mathrm{ref}$ is smoothly linked to the solar regime and will likely also give plausible results for stars with \dnu\ outside the range of our eclipsing benchmark stars of about 3 -- 7\mh . While anchoring to the solar case practically ensures reasonable results for stars with higher \dnu , for lower \dnu\ an extrapolation is involved, which makes the results more questionable. In Sec.\,\ref{Sec:ClusterStars} we, however, verify Eq.\,\ref{eq:nonlinear_dnu} for stars with \dnu\ down to about 0.9\mh. 

For the asymptotic large separation we do not find a gradually decreasing reference value. In fact, the scatter is much larger than for \dc\ and \dcor\ and we can only establish an average reference value of \dnu$_\mathrm{as,RGB} = 142.2\pm1.6$\mh .

\subsection{Re-determined seismic fundamental parameters}
Above we have determined RGB-calibrated reference values for the classical linear scaling relations (Eq.\,\ref{eq:dnu} and \ref{eq:numax}) and established new nonlinear scaling relations (Eq.\,\ref{eq:numDynRelation} and \ref{eq:nonlinear_dnu}). We now apply these scalings to re-detemine the seismic fundamtental parameters of our sample stars.

In a first step, we use the RGB-calibrated average reference values $\nu_\mathrm{max,RGB}$ and \dnu$_\mathrm{c,RGB}$, \dnu$_\mathrm{cor,RGB}$, and \dnu$_\mathrm{as,RGB}$ and find the systematic deviations between seismic (determined from Eq.\,\ref{eq:dnu} and \ref{eq:numax}) and dynamic parameters to largely disappear (see Tab.\,\ref{tab:RMsystematic}). This is not surprising since the new reference values are chosen to minimise the deviation between seismic and dynamic measurements. Note that \cite{themessl2016} follow a similar approach. Interesting is that the scatter of the deviations is similar to those of the solar-calibrated fundamental parameters. This means that the random uncertainties are dominated by the errors in the seismic and dynamic observables, and that the relatively large uncertainties in the reference values, compared to solar, have only marginal effects. 

This changes if we use the nonlinear scalings (Eq.\,\ref{eq:numDynRelation} and \ref{eq:nonlinear_dnu}) for \num\ and \dnu$_\mathrm{c}$ and \dnu$_\mathrm{cor}$ (see Fig.\,\ref{fig:RMcalib} and Tab.\,\ref{tab:RMsystematic}). This approach gives seismic fundamental parameters that come close to their dynamic counterpart with marginal systematic deviations and the lowest random uncertainties of all investigated combinations of observed parameters and reference values.

In summary, the 4 year-long \kep\ observations of red giants allow us to seismically constrain the mass and radius of RGB stars with typical internal uncertainties of only a few per cent. While the classical solar-calibrated approach results in significant systematic uncertainties, our new approach with nonlinear scalings largely eliminates this systematics while leaving the random uncertainties basically untouched (see Tab.\,\ref{tab:RMcalib}). We note that the statistics provided in this analysis is based on a sample of only six stars. The uncertainties that we give are therefore only tentative. 

\begin{figure}[t]
	\begin{center}
	\includegraphics[width=0.5\textwidth]{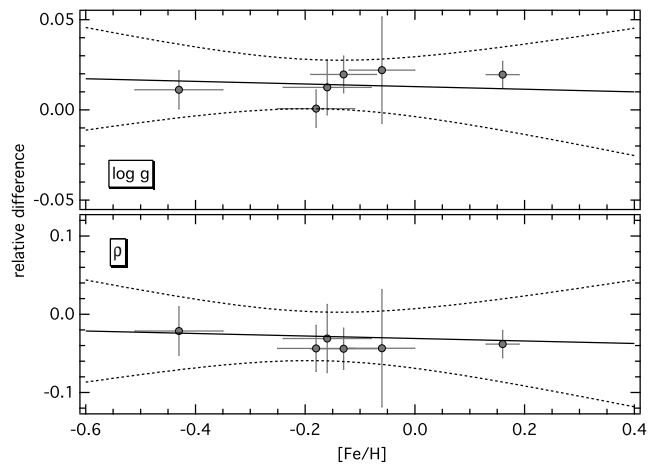}
	\caption{Relative difference between the seismic and dynamic parameters as a function of stellar metallicity for the solar-calibrated surface gravity (top) and mean stellar density scaled from $\Delta\nu_\mathrm{cor}$ (bottom). Solid and dotted lines indicate a linear fit and the corresponding 95\% confidence intervals, respectively.} 
	\label{fig:FeH} 
	\end{center} 
\end{figure}

\begin{figure}[t]
	\begin{center}
	\includegraphics[width=0.5\textwidth]{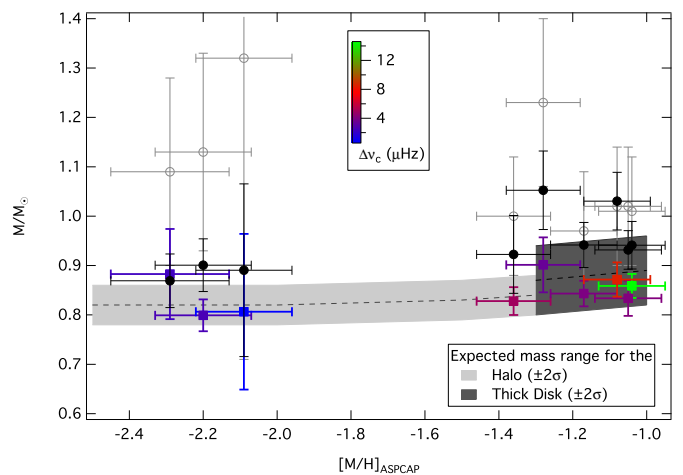}
	\caption{Seismic masses for a sample of halo and thick-disk stars as a function of stellar metallicity. The open grey  symbols show the original values taken from \cite{epstein2014}. Black symbols correspond to masses from the classical scaling relations as done by \cite{epstein2014}, but based on $\Delta\nu_\mathrm{c}$ determined from the full \kep\ data sets. Coloured symbols show the masses based on our re-calibrated scaling relations, where the colour indicates the observed large frequency separation $\Delta\nu_\mathrm{c}$. The grey-shaded bars indicate the expected mass based on the high age and low metallicity constraints for these stars \citep{epstein2014}.
    } 
	\label{fig:HaloStars} 
	\end{center} 
\end{figure}

\subsection{Stellar metallicity and the seismic scalings}

It has been suggested that at least the $\Delta\nu - \bar\rho$ scaling should in some sense be sensitive to stellar metallicity \citep[e.g.,][]{white2011,epstein2014,guggenberger2016,sharma2016}. However, we find no evidence for such a dependency as is shown in Fig.\,\ref{fig:FeH}, where we plot the relative difference between the solar-calibrated seismic and dynamic $g$ and $\rho$ as a function of stellar metallicity. Even though there is a small trend for both quantities, it remains statistically insignificant. Our sample is too small and not accurate enough to proof or disproof a metallicity dependency. 

On the other hand, \cite{epstein2014} found that the (solar-calibrated) seismic scaling relations overestimate the stellar mass by about 17\% for a sample of halo and thick-disk stars, for which the mass and age are well-constrained by astrophysical priors. They attribute the discrepancy to a metallicity effect. 

As their seismic analysis was not based on the full \kep\ time series we re-analysed their sample with the above described methods. Due to the low \dnu\ and/or high noise level in the data, we can thereby determine $\Delta\nu_\mathrm{cor}$ for only about half of the sample stars. To be consistent, we restrict the further analysis to $\Delta\nu_\mathrm{c}$ but note that the results are similar when based on $\Delta\nu_\mathrm{cor}$ (if available). Using the new and significantly more accurate global seismic parameters we derive classical solar-calibrated masses that are in much better agreement with the expectations (see Fig.\,\ref{fig:HaloStars}). Even though the seismic masses are still systematically too high, the discrepancy reduces by about a factor of two. However, if we apply the nonlinear scaling relations for \num\ and $\Delta\nu_\mathrm{c}$ we find seismic masses that are just within the expectations for halo and thick-disc stars \citep[see][]{epstein2014}. The claimed ``metallicity effect'' can therefore be explained equally well by a necessity for nonlinear scalings, or in order words by the inaccuracy of the classical linear scaling for such luminous stars. Even though there are other uncertainties in this analysis, the now good agreement for the halo stars supports that our re-calibrated scaling relations are better for estimating mass than the classical relations, and that the dynamic parameters on which our relations are based, are correct.

\begin{table}[t]
\begin{tiny}
\begin{center}
\caption{Non-seismic and seismic radius of the bright RGB star HD\,185351 observed by \kep\ in short-cadence mode. 
\label{tab:hd185351}}
\begin{tabular}{lccccc}
\hline\hline
\noalign{\smallskip}
 & non-seismic& \multicolumn{3}{c}{seismic} \\
 & Johnson et al. & Sharma et al. & \multicolumn{2}{c}{this work}\\
\noalign{\smallskip}
\hline
\noalign{\smallskip}
\num\ & & & 229.8$\pm$6.0	& 230.9$\pm$3.6 \\
\dnu\ & & & 15.4$\pm$0.2 & 15.51$\pm$0.27 \\
\noalign{\smallskip}
\hline
\noalign{\smallskip}
R/R\sun &   4.97$\pm$0.07	& 5.23$\pm$0.2	& 5.08$\pm$0.14	& 5.06$\pm$0.16\\
\noalign{\smallskip}
\hline
\noalign{\smallskip}
\end{tabular}
\tablefoot{The global seismic parameters (\num\ and \dnu\ in \mh) listed in the last two columns correspond to the original values from \cite{Johnson2014} and our own estimates (\dc ), respectively.}
\end{center}
\end{tiny}
\end{table}

\subsection{The benchmark case HD\,185351}
Since \kep\ red giants usually lack accurate non-seismic parameters, it is difficult to independently verify our newly established nonlinear scalings. A rare example, where this is nonetheless possible is the bright ($V=5.18$) RGB star HD\,185351 observed with \kep , for which \cite{Johnson2014} determined a precise radius from interferometry and spectral energy distribution fitting (see Tab.\,\ref{tab:hd185351}). \cite{sharma2016} used the star as a benchmark case in their analysis as well and find a seismic radius that overestimates the interferometric values by about 5$\pm$4\%. Our nonlinear scalings (for \num\ and $\Delta\nu_\mathrm{c}$) reduce the discrepancy to about 2.1$\pm$3.2\% if we use the original global seismic parameters from \cite{Johnson2014} as input. Even though the star is very bright the \textit{Kepler} observations of HD\,185351 are quite noisy and reveal significant modes only in the central three to four radial orders of the pulsation power excess. We can therefore only determine $\Delta\nu_\mathrm{c}$. Based on this the radius discrepancy reduces to 1.7$\pm$3.6\%. We further note that our nonlinear scalings give a mass of $1.72\pm0.14$\,M\sun , which is in good agreement with the value of 1.6\,M\sun\ found by \cite{Hjorringgaard2017} based on a detailed seismic analysis.

\section{Cluster stars} \label{Sec:ClusterStars}
Using the dynamic fundamental parameters of a sample of eSB2  binaries we have established new nonlinear seismic scaling relations. The calibration sample is, however, quite small covering only a small range of evolutionary stages ($\sim$7.6 -- 14\,$R$\sun) and is limited to RGB stars, burning hydrogen in a shell around an inert helium core. In order to test our scaling relations beyond this limitation, we need a larger sample of stars covering various evolutionary stages, including stars in the red-clump (RC) phase, of quiescent core-helium burning, with homogenous seismic measurements and additional accurate non-seismic constraints. The \kep\ observations of the red giants in the two well-studied open clusters NGC\,6791 and NGC\,6819 form such a sample. 

\subsection{NGC 6791}
\na\ is one of the oldest, and at the same time, most metal-rich open clusters known \citep[e.g.,][]{Carretta2007}. It is a well-populated cluster, with a significant number of stars at basically all stages of evolution from cool main-sequence stars to white dwarfs \citep[e.g.,][]{King2005,Bedin2005} including many variable stars \citep[e.g.,][]{Mochejska2002,deMarchi2007}. Despite the large number of studies there is little agreement on its age, ranging 7--12\,Gyr, and metallicity because of correlated uncertainties in distance, reddening, and metallicity, with the latter ranging from [Fe/H] = +0.29 \citep{Brogaard2011} to +0.45 \citep{Anthony-Twarog2007}. It seems likely that the differences are mainly caused by differences in the adopted reddening, with typical values of either $E_{(B-V)}$ = 0.09 \citep{Stetson2003} or 0.15 \citep[an average of literature determinations,][]{Carretta2007}. \cite{An2015} carefully calibrated the clusters' colour-magnitude diagram (CMD) and report a true distance modulus (DM) of $(m-M)_0=13.04\pm0.08$\,mag and an age of $9.5\pm0.3$\,Gyr from isochrone fitting, where the latter uncertainties do not cover systematic uncertainties in the isochrones. A detailed literature overview can be found in \cite{wu2014b}. 

There are several studies available \citep{basu2011,miglio2012,wu2014a,wu2014b} that aimed to constrain the RGB mass and true DM from \kep\ photometry of red giants in \na. Even though these studies are different in detail they converge to M$_\mathrm{RGB}=1.23\pm0.02$\,M\sun\ and $(m-M)_0=13.1\pm0.1$\,mag. \cite{Brogaard2012}, on the other hand, suggested a slightly lower RGB mass of $1.15\pm0.02$\,M\sun\ based on isochrone fitting to the mass and radius of several eclipsing binaries on the main sequence of \na. \cite{miglio2012} furthermore found a small but significant difference between the average mass of RGB and RC stars of $\Delta\bar M = 0.09\pm0.05$\,M\sun, indicating only moderate mass-loss for this metal-rich cluster.

\begin{figure}[t]
	\begin{center}
	\includegraphics[width=0.5\textwidth]{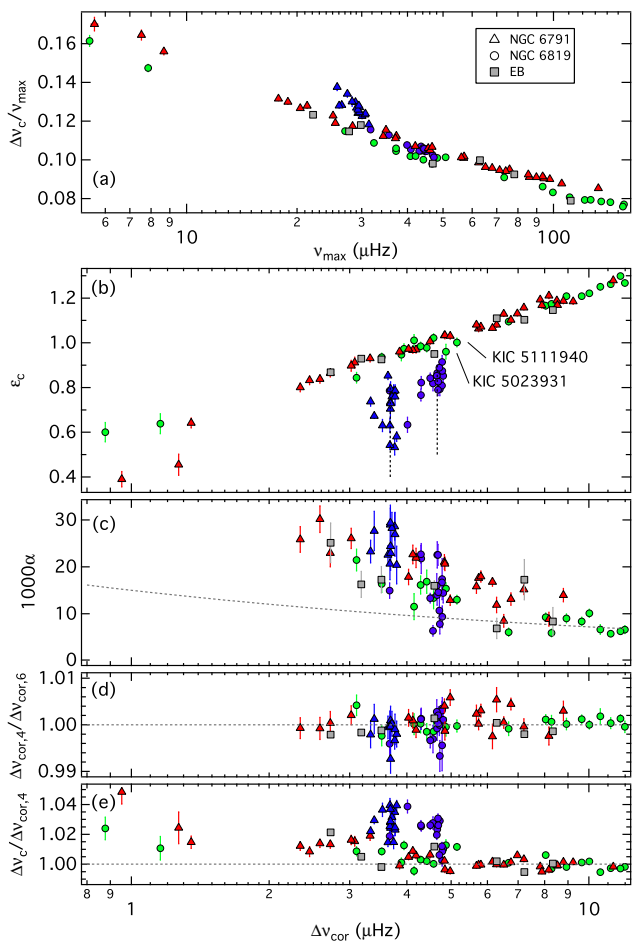}
	\caption{Global seismic parameters for \na\ and \nb , where the RGB stars are indicated by green circles and red triangles, repsectively, and the RC stars of both clusters by blue sysmbols. Grey-filled squares show the EBs from Sec.\,\ref{sec:eb}. Panel (a) shows the ratio of $\Delta\nu_{c}$ and \num\ and panel (b) the phase shift $\epsilon_c$ of the central radial mode. The vertical dashed lines indicate the median large separations of RC stars in the two clusters. Panel (c) gives the curvature term $\alpha$. Panel (d) shows the ratio of $\Delta\nu_\mathrm{cor}$ computed in different ways (Eq.\,\ref{eq:asymrel0} and \ref{eq:glitch}) and panel (e) the ratio of central large separation and $\Delta\nu_\mathrm{cor}$ computed according to Eq.\,\ref{eq:asymrel0}. The dashed lines in panel (c) and (d,e) indicate the scaled curvature term from \cite{mosser2013} and unity, respectively.} 
	\label{fig:seispar} 
	\end{center} 
\end{figure}

\subsection{NGC 6819}
\nb\ is a very rich open cluster with roughly solar metallicity
of [Fe/H] = $+0.09\pm0.03$, reddening $E_{(B-V)} = 0.14\pm0.04$ \citep{Bragaglia2001}, and age of about 2.5\,Gyr \citep[isochrone fitting in the CMD; e.g.,][]{Kalirai2004}. Since there is reasonable agreement on the metallicity and reddening of \nb\ it has been used to study other phenomena, such as the initial-final mass relation using the clusters' white dwarf population \citep[e.g.,][]{Kalirai2008}. \cite{Jeffries2013} reported, however, that the detached eclipsing binary KIC\,5113053, located near the cluster turnoff, is significantly older than the commonly adopted cluster age. They fit the mass and radius of the binary components with isochrones from various stellar evolution codes and find ages ranging from 2.9 to 3.7\,Gyr while the same isochrones give values between 2.1 and 2.4\,Gyr for fits in the CMD. This clearly indicates that the age determination of stellar clusters is still problematic and obviously quite model-dependent. As for \na , \cite{basu2011} used the \kep\ photometry to determine the RGB mass and true DM of \nb\ to M$_\mathrm{RGB}=1.68\pm0.03$\,M\sun\ and $(m-M)_0=11.85\pm0.05$\,mag, which is again compatible with the estimates of $1.75\pm0.03$\,M\sun\ and $11.88\pm0.14$\,mag reported by \cite{wu2014a,wu2014b}. \cite{miglio2012} found no mass-loss on the clusters' giant branch but a smaller RGB mass of $1.61\pm0.04$\,M\sun, which is supported by \cite{Sandquist2013} who expected M$_\mathrm{RGB}$ to be about 8\% smaller than reported based on isochrone fitting to the mass and radius of EBs on the clusters' main sequence.  

\subsection{Input parameters}
\subsubsection{Global seismic parameters \num\ and $\Delta\nu_\mathrm{cor}$}
\begin{figure*}[t]
	\begin{center}
	\includegraphics[width=1.0\textwidth]{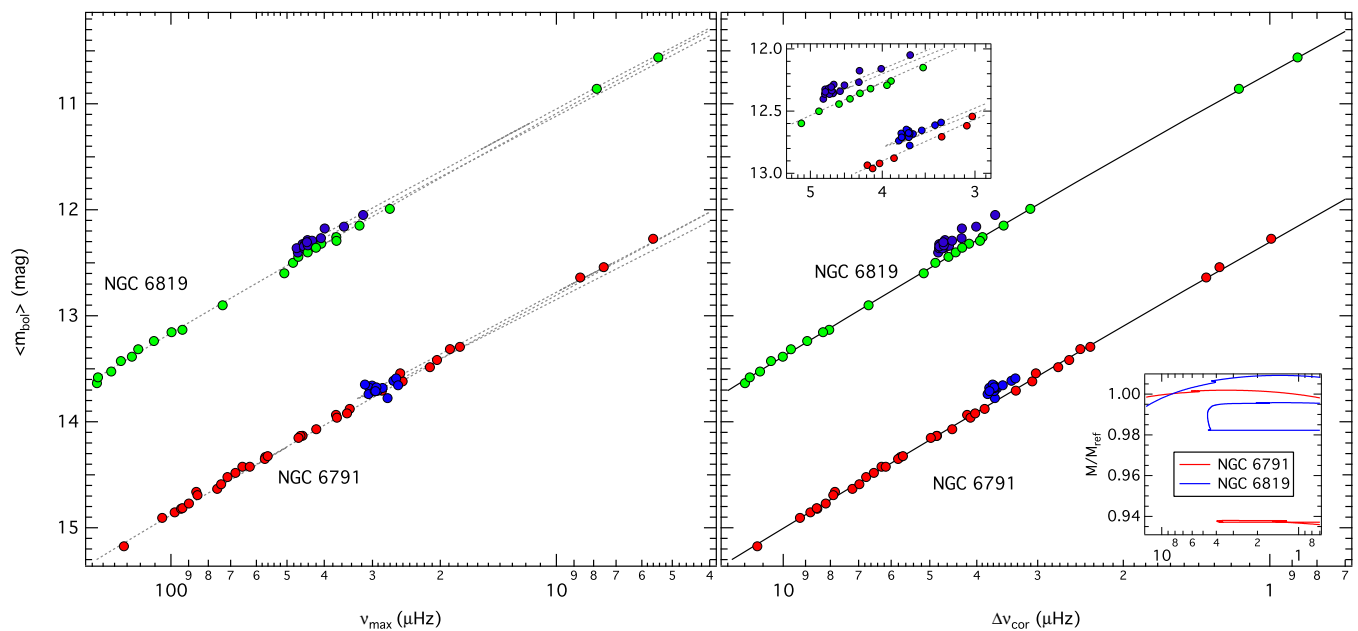}
	\caption{Average apparent bolometric magnitude as a function of \num\ (left panel) and $\Delta\nu_\mathrm{cor}$ (right panel) for our sample of RGB (red and green dots) and RC (blue dots) stars in \na\ and \nb , respectively. The dotted lines in the left panel (and top inset in the right panel) correspond to \textit{BaSTI} isochrones (Z=0.03 and 8.0\,Gyr for \na\ and Z=0.02 and 2.5\,Gyr for \nb ) arbitrarily shifted to match the observations. Note that the loops in the isochrones correspond to the various phases of the RGB evolution. Solid lines in the right panel indicate polynomial fits to the RGB stars. The bottom inset gives the mass of the \textit{BaSTI} isochrones normalised to the mass at $\nu_\mathrm{max}=100$\mh .} 
	\label{fig:cmd} 
	\end{center} 
\end{figure*}

A prerequisite for the subsequent analysis is that the stars of interest are members of the clusters. As a starting point we use the list of ``seismic members'' from \cite{stello2011} from which we exclude misclassified stars, suspected binaries, and stars with other ``peculiarities'' \citep[e.g.,][]{Corsaro2012,Bellamy2015}. The about 1420 days-long \kep\ time series of the remaining cluster members are processed in the same way as the G16 sample (see Sec.\,\ref{sec:seispar}). In a first step, we extract \num\ and $\Delta\nu_c$ from the power density spectra of all stars and find values with sufficient accuracy for 51 and 39 stars in \na\ and \nb, respectively. They are shown in Fig.\,\ref{fig:seispar}a, where they form two slightly different sequences indicating the different RGB masses of the clusters \citep[e.g.,][]{Hekker2011}. To disentangle RC stars from RGB star, we compute the phase shift of the central radial mode ($\epsilon_c$; see Fig.\,\ref{fig:seispar}b). According to \cite{kal12} and \cite{ChristensenDalsgaard2014}, $\epsilon_c$ is expected to be significantly smaller for RC stars than for RGB star at a given $\Delta\nu$. Unlike claimed by \cite{handberg2017} for stars in \nb, we can in fact unambiguously identify the evolutionary stage of almost all stars with this method. Only for two stars in \nb\ (KIC\,5023931 and KIC\,5111940) the result remains uncertain. A visual inspection of their dipole mode spectrum does, however, reveal that both stars are very likely RGB stars (which is confirmed by the stars positions in the seismic colour-magnitude diagram; see Sec.\,\ref{sec:clusterlum} and Fig.\,\ref{fig:cmd}). The two stars are not part of other RGB-RC classification analysis \citep[e.g.,][]{vrard2016,Elsworth2017,Hon2017}. 

Subsequently, we try to peakbag the radial mode frequencies for all stars in order to determine the curvature- and glitch-corrected large separation \dcor\ (see, Sec.\,\ref{sec:asymDnu}). However, we are only able to measure \dcor\ with sufficient accuracy for 32 out of 51 \na\ members and 34 out of 39 \nb\ members. The corresponding curvature parameters $\alpha$ are given in Fig.\,\ref{fig:seispar}c, showing a large intrinsic scatter between the two clusters, presumably due to a mass and/or metallicity dependency. We also see the overall trend significantly deviates from that provided by \cite{mosser2013}. In order to avoid losing a significant number of stars for the subsequent analysis, especially at low $\Delta\nu$, we use a different method to determine $\Delta\nu_c$. 

\begin{figure}[t]
	\begin{center}
	\includegraphics[width=0.5\textwidth]{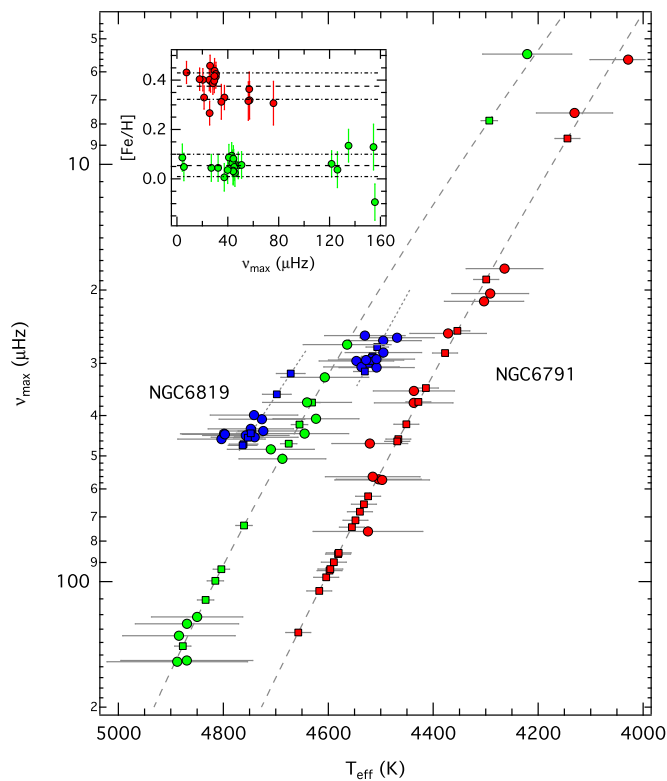}
	\caption{Peak frequency vs. effective temperature for our sample of RGB (red and green symbols) and RC (blue symbols) stars in NGC6791 and NGC6819, respectively. Circles and squares indicate stars with the effective temperature taken from the APOKASC Catalogue and derived from polynomial fits (grey dashed and dotted lines), respectively. Horizontal grey lines represent the uncertainties in \teff. The inset shows the metallicity measurements from the APOKASC Catalogue along with the average values (dashed lines) and their uncertainties (dashed-dotted lines).} 
	\label{fig:temp} 
	\end{center} 
\end{figure}

Our original approach to extract $\Delta\nu_c$ is based on fitting a sequence of Lorentzian profiles to the power density spectrum that cover the $l=0$ to 2 modes in the three radial orders around the central radial mode. We now extend the range to five radial orders and add a curvature term to the parameterisation of the radial mode frequencies according to Eq.\,\ref{eq:asymrel0}, rather than treating them as equidistant as in the original approach. To test if the resulting $\Delta\nu_\mathrm{cor}$ are sufficient for the subsequent analysis we compare them to \dcor\ determined from Eq.\,\ref{eq:glitch} for the stars for which we could successfully peakbag the radial modes in Fig.\,\ref{fig:seispar}d.  We find that the individual values agree with each other within $\sim$0.30\% on average, which is not much larger than the average uncertainty of their ratio of $\sim$0.25\%. This indicates that a remaining glitch contribution, to \dcor\ as determined from Eq.\,\ref{eq:asymrel0}, is quite small and essentially covered by the uncertainties delivered by \textsc{MultiNest}. We therefore assume \dcor\ from Eq.\,\ref{eq:asymrel0} to be precise enough for the subsequent analysis.

Finally, we show the ratio $\Delta\nu_c / \Delta\nu_\mathrm{cor}$ in Fig.\,\ref{fig:seispar}e. This indicates that $\Delta\nu_c$ is a good representation of the ``true'' $\Delta\nu$ for the least evolved RGB stars (large $\Delta\nu$) but the curvature and glitch contribution systematically increases for more evolved RGB stars. For RC stars, $\Delta\nu_c$ includes a significant contribution from curvature and glitches that is much larger than the observational uncertainties, which needs to be corrected if one is interested in the real large separation. More details about the curvature and glitch signal will be presented in a separate study (Kallinger et al., in preparation).   

\subsubsection{Effective temperatures}
The cluster stars' effective temperatures are taken from the APOKASC Catalogue \citep[DR13;][]{pin2014}. Unfortunately, only about half of our sample stars are listed in the catalog. For stars with near-equal mass, however, we can expect a tight relation between \num\ and effective temperature. As is shown in Fig.\,\ref{fig:temp}, there is a clear relation, so that we can fit \teff\ as a function of $\log{\nu_\mathrm{max}}$, for which we use first (RGB) and second order (RC) polynomial fits. Effective temperatures for the remaining stars are then determined from the fits according to their \num. Fig.\,\ref{fig:temp} also shows that the uncertainties listed in the APOKASC Catalogue (typically about 80-90\,K) are significantly larger than the scatter around the fits. This is because the APOKASC errors are supposed to also cover systematic uncertainties. In our analysis we are more interested in random errors and therefore adopt uncertainties for all stars that correspond to the rms scatter around the fits, which is $\pm$24 (RGB) and $\pm$21\,K (RC) for stars in NGC 6791, and $\pm$16 (RGB) and $\pm$27\,K (RC) for stars in NGC 6819). Hence we assume any systematic \teff\ errors are the same for all red giants. The average metallicities ([M/H]) for the stars listed in the catalog are $+0.38\pm0.05$ for NGC 6791 and $+0.05\pm0.04$ for NGC 6819, which is in good agreement with previous measurements. 

\subsubsection{Stellar luminosities} \label{sec:clusterlum}
Stellar luminosities are usually determined by the apparent magnitude of a star and its distance but to derive distances is a difficult task. Several classical studies of the two clusters revealed distance moduli that diverge by up to half a magnitude. The situation improved with the availability of precise global oscillation parameters from \kep, which improved the uncertainties to about 0.1--0.15\,mag. This implies that the luminosity of individual cluster stars can not be determined to better than about 10-12\%, which makes them impractical for our test of the seismic scaling relations.

However, the fact that the cluster members are equally distant\footnote{Assuming that the apparent cluster diameters of about 16 and 5\,arcminutes are also a good estimate for the clusters extention along the line of sight, the distance dispersion is for both clusters on the order of a few thousandths.} and that their light undergoes approximately the same reddening and extinction enables us to derive relative luminosities that depend only on the differences between the apparent bolometric magnitudes ($m_\mathrm{bol}$) of the individual stars.  Red giants are brightest in the near-infrared so that the 2MASS $JHK$ photometry \citep{skr2006} is well suited to determine accurate $m_\mathrm{bol}$. We bolometrically correct the individual 2MASS magnitudes by interpolating $T_\mathrm{eff}$, $\log g$, and metallicity in the tables of \citet{bonatto2004}, where the surface gravity is estimated from $g \propto \nu_\mathrm{max} \sqrt{T_\mathrm{eff}}$. The uncertainties of all parameters are fully taken into account and the final apparent bolometric magnitude (and its uncertainty) is computed from the average of the three individual values.

\begin{figure*}[t]
	\begin{center}
	\includegraphics[width=1.0\textwidth]{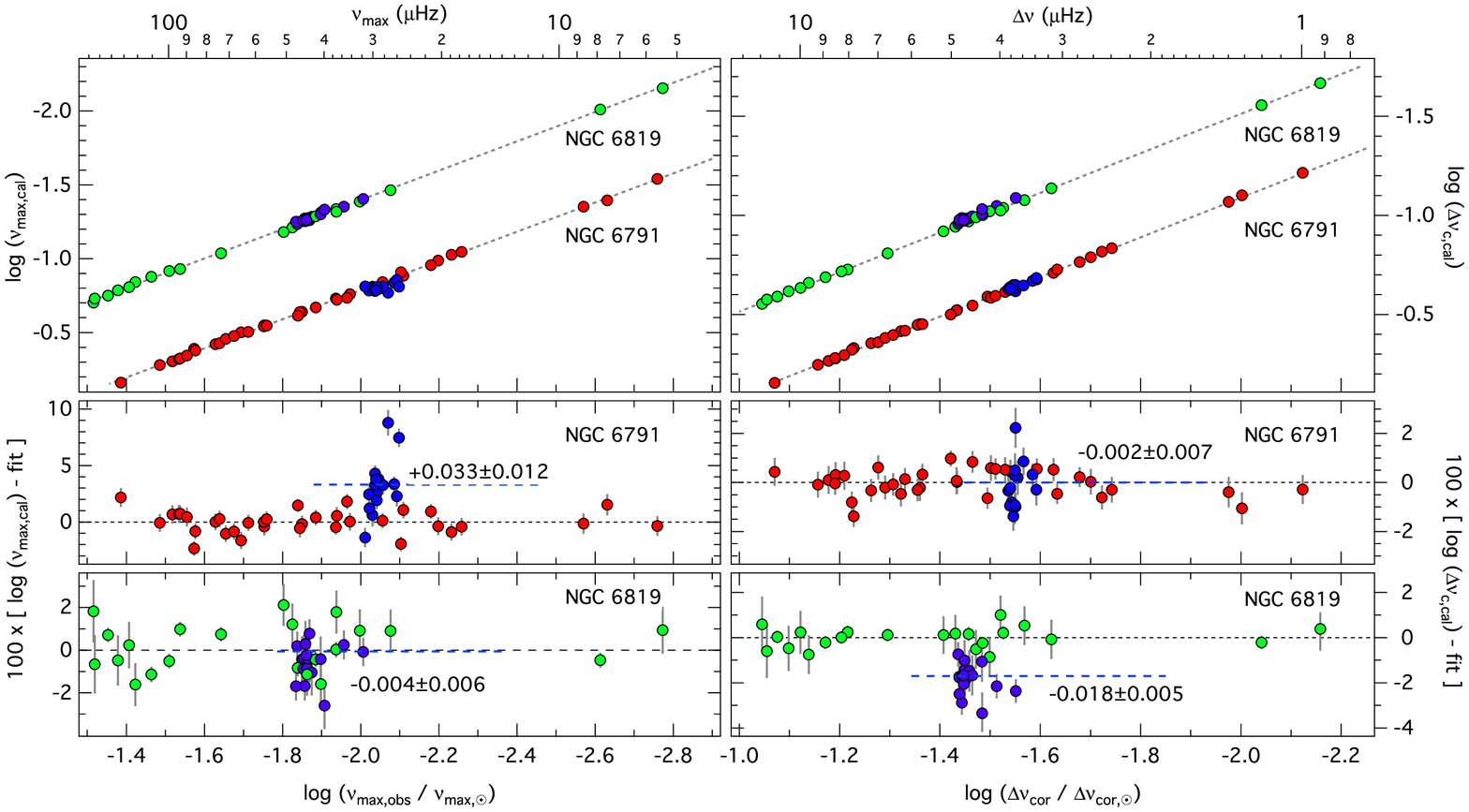}
	\caption{\textit{Top:} The from $m_\mathrm{bol}$ and $T_\mathrm{eff}$ scaled $\nu_\mathrm{max,cal}$ (left) and $\Delta\nu_\mathrm{cor,cal}$ (right) as a function of the observed parameters for the sample of RGB (red and green dots) and RC (blue dots) in \na\ and \nb , respectively. The dashed lines indicate linear fits (in log-log scale) to the RGB stars, which correspond to powerlaw fits in linear scale. Note that the panels are plotted upside down so that bright stars are at the top. \textit{Middle and bottom:} Residuals to the fits. The blue dashed lines indicate the average offset of RC stars with the mean values and rms scatter given.} 
	\label{fig:ClusterSeis} 
	\end{center} 
\end{figure*}

In Fig.\,\ref{fig:cmd} we show $m_\mathrm{bol}$ as a function of \num\ and \dcor. These seismic colour-magnitude diagrams (sCMD) are a powerful tool for the characterisation of cluster stars. All stars that are single members are supposed to follow a distinct sequence (indicated here by BaSTI isochrones; \citealt{piet2004}). If a star, assumed single, falls below the sequence clearly indicates a background star. If a stars falls above the sequence it is either a foreground star or an unresolved binary combining the light of more than one object. In the present case the membership analysis \citep{stello2011} was already done carefully enough so that all stars in our sample are clearly cluster members. Additionally, the sCMD does even allow us to distinguish between RGB and RC stars. The effect is slightly smeared out when taking \num\ as a reference but for \dcor\ RGB and RC stars are clearly separated. This is not surprising because RC stars are slightly hotter (see Fig.\,\ref{fig:temp}) and therefore more luminous than RGB stars with the same \dcor. We note that the actual identification used here is based on the phase shift of the central radial mode \citep{kal12} and therefore entirely independent. 

Linear fits in the $m_\mathrm{bol}-\log{\Delta \nu_\mathrm{cor}}$ plane to the RGB stars reveals that the apparent bolometric magnitudes are accurate to about 0.02\,mag for both clusters. This implies that we can constrain the relative luminosities to about 1.8\%, which is at least five times more accurate than the absolute luminosities.

\subsubsection{Stellar masses}
In a cluster with all stars having the same age and primordial chemical composition there is a tight relation between the initial mass of a star and its evolutionary stage, with more massive stars being further evolved. On the giant branch, however, stars evolve fast, compared to main-sequence stars, so that high-luminosity RGB stars are expected to be only slightly more massive than low-luminosity RGB stars. This is shown in the inset of Fig.\,\ref{fig:cmd}, where we indicate how mass is expected to change along the giant branch. According to \textit{BaSTI} isochrones (which include mass loss using the \cite{Reimers1975} formulation with the free parameter $\eta$ set to 0.2) the RGB star mass in \nb\ is supposed to increase by about 1\% from the least evolved to the most evolved star in our sample. Such high sensitivity to mass is also observed in the binary system KIC\,9163796, which consists of a sub giant and a red giant with a mass ratio of $1.01\pm0.01$ \citep{beck2017}. For \na\ the mass dispersion is even smaller as the evolutionary mass increase is balanced by mass loss (based on $\eta=0.2$). It is therefore save to assume that all RGB stars in our sample have approximately the same mass, $M_\mathrm{RGB}$.

\subsection{Comparision of observed and computed global seismic parameters for RGB stars}

With independent constraints for the mass, effective temperature, and luminosity in hand, we can now compare the global seismic parameters estimated from scaling relations to the actual observed ones.
\subsubsection{Testing the \num$-$scaling}
Given Eq.\,\ref{eq:numax} and $L\propto R^2T^4$ the peak frequency \num\ scales as 
\begin{equation} \label{eq:numax_lum}
\nu_\mathrm{max} \propto M\,R^{-2}\,T_\mathrm{eff}^{-1/2} \propto M\,L^{-1}\,T_\mathrm{eff}^{7/2}. 
\end{equation}
The solar-scaled luminosity can be determined according to
\begin{equation}
L = 79.829\cdot 10^{-0.4(m_\mathrm{bol}-DM)},
\end{equation}
where the factor 79.829 comes from the absolute bolometric magnitude of the Sun, $M_{\mathrm{bol},\sun}$ = 4.7554\,mag \citep{kopp2011}.
By combining the above equations, we can write the logarithmic peak frequency, $\log\nu_\mathrm{max,cal}$, as
\begin{equation} \label{eq:numax_cal}
\log{\nu_\mathrm{max,cal}} = -1.902+\log{M} - 0.4\, DM + 3.5 \log{T_\mathrm{eff}} + 0.4\,m_\mathrm{bol},
\end{equation}
where $\nu_\mathrm{max,cal}$, $M$, and $T_\mathrm{eff}$ are in solar units. This description of the peak frequency is independent of any seismic constraint and corresponds to the right-hand side of Eq.\,\ref{eq:numax_lum}. 

If the \num --scaling would be correct, $\nu_\mathrm{max,cal}$ would linearly scale with the observed peak frequency. This can be tested by fitting a function:
\begin{equation} \label{eq:numaxfit}
\log{\nu_\mathrm{max,cal}} = k \cdot \log{(\nu_\mathrm{max,obs}/\nu_\mathrm{max,\odot})} + d,
\end{equation}
to the RGB stars of both clusters, where the slope $k$ would need to be equal to one if the scaling is linear. We show $\nu_\mathrm{max,cal}$ as a function of the observed peak frequency along with the best fit in Fig.\,\ref{fig:ClusterSeis}.  We find a best fit of $k=0.989\pm0.004$ for \na\ and $0.995\pm0.004$ for \nb. A possible explanation for $k\neq1$ are potential systematic uncertainties in the adopted effective temperatures. However, it turns out that we need to subtract about 400 (\na) and 250\,K (\nb), to obtain slopes that are consistent with one. Such large systematics do not seem very likely. A reconciliation can be obtained by assuming a nonlinear scaling of the form $g/\sqrt{T_\mathrm{eff}} \propto (\nu_\mathrm{max,obs}/\nu_\mathrm{max,\odot})^{\kappa}$, where $\kappa = 1/k$. The slopes from above translate into exponents of $1.010\pm0.004$ and $1.005\pm0.004$, which are consistent with the value of $1.008\pm0.002$ determined from the eSB2 stars in Sec.\,\ref{sec:newrefs}. Combining the three independent cluster and eSB2 measurements gives an average value of $\kappa=1.0075\pm0.0021$. We also tested higher-order polynomial fits instead of Eq.\,\ref{eq:numaxfit} but find them to be statistically insignificant. 

\subsubsection{Testing the $\Delta\nu_\mathrm{cor}-$scaling}
As for the peak frequency, we can express the large separation as a function of $M$, $L$, and $T_\mathrm{eff}$ according to,
\begin{equation}
\Delta\nu_\mathrm{cor} \propto M^{1/2} R^{-3/2} \propto M^{1/2} L^{-3/4} T_\mathrm{eff}^{3},
\end{equation}
which translates into a logarithmic description as,
\begin{equation} \label{eq:dnu_cal}
\log{\Delta\nu_\mathrm{cor,cal}}=-1.423 + 0.5\log{M} -0.3DM + 3\log{T_\mathrm{eff}} + 0.3m_\mathrm{bol}.
\end{equation}
We then again test the linearity of the $\Delta\nu_\mathrm{cor}-$scaling by fitting the function:
\begin{equation}	\label{eq:dnufit}
\log{\Delta\nu_\mathrm{cor,cal}} = k \cdot \log{(\Delta\nu_\mathrm{cor,obs}/\Delta\nu_\mathrm{cor,\odot})} +d, 
\end{equation} 
to the RGB stars in \na\ and \nb . Expectedly, we find $k$ with $0.981\pm0.003$ and $0.980\pm0.004$, which differs significantly from unity in both clusters. The excellent agreement between the two slopes indicates again that the \dnu$-$scaling is not affected by metallicity; at least not beyond 0.1$-$0.3\% in the \teff\ range for which \kep\ has observed red giants in the two clusters. 

Unlike for the exponent in the \num$-$scaling, the slope $k$ in Eq.\,\ref{eq:dnufit} cannot be directly translated into the parameter $\gamma$ of the nonlinear $\Delta\nu-$scaling presented in Eq.\,\ref{eq:nonlinear_dnu}. It is, however, evident that the cluster stars do also support a nonlinear scaling and replacing $\Delta\nu_\mathrm{cor,\odot}$ in Eq.\,\ref{eq:dnufit} by $\Delta\nu_\mathrm{cor,\odot}[1-\gamma \log^2{(\Delta\nu_\mathrm{cor,obs}/\Delta\nu_\mathrm{cor,\odot})}]$ basically linearises the $\Delta\nu-$scaling (meaning $k\simeq1$) if we adopt the original value of $\gamma=0.0085$.

\subsection{Extending the scalings to RC stars}

So far we have only considered RGB stars and from the middle and bottom panels of Fig.\,\ref{fig:ClusterSeis} it is evident that the scaling relations cannot be directly applied to RC stars. For the peak frequency, subtracting the linear fits from all stars results in a median underestimation of $\log{\nu_\mathrm{max}}$ of $0.033\pm0.012$ for the clump stars in \na, while no systematic deviation is found for the RC stars in \nb. The obvious explanation for this is mass loss in \na . The offsets $d$ in the fits represent the RGB mass and distance moduli of the two clusters. Since the distance is practically the same for all cluster members, a systematic difference between the fit and RC stars can only mean that the mass is different between RGB and RC stars. In fact, we can estimate that $M_\mathrm{RC}/M_\mathrm{RGB}=10^{-0.033\pm0.012}$. Or, with other words, the RC stars in \na\ have lost $7.5\pm2.5$\% of their mass between the late RGB and clump phase, which is in agreement with, but determined more accurately than, the findings of \cite{miglio2012}. 

Subtracting this mass loss for clump stars in Eq.\,\ref{eq:dnu_cal} leads to an median overestimation of $\log{\Delta\nu_\mathrm{cor}}$ of $-0.018\pm0.009$ for clump stars in \na, which is in agreement with the value of $-0.018\pm0.005$ found for \nb. Because we find the same systematic difference in both cluster it indicates that the $\Delta\nu-$scaling should be different for RC stars. From the average value we can estimate a correction factor of $1.04\pm0.01$ (i.e., $10^{0.018\pm0.05}$) for the large separation of clump stars, which is slightly larger than the value suggested by \cite{miglio2012}. From this we can finally formulate our new nonlinear seismic scaling relations as:
\begin{equation}	\label{eq:final_numax}
\frac{g}{\sqrt{T_\mathrm{eff}}} = \left (\frac{\nu_\mathrm{max}}{\nu_\mathrm{max,\odot}} \right)^{1.0075\pm0.0021}
\end{equation}
and
\begin{equation}	\label{eq:final_dnu}
\sqrt{\bar\rho} = \frac{\Delta\nu_\mathrm{cor} }{ \Delta\nu_\mathrm{cor,\odot}} \left [\eta - (0.0085\pm0.0025) \cdot \log^2 (\Delta\nu_\mathrm{cor} / \Delta\nu_\mathrm{cor,\odot}) \right ]^{-1},
\end{equation}
where $g$, $T_\mathrm{eff}$, and $\bar\rho$ are in solar units, $\nu_\mathrm{max,\odot}=3140\pm5$\mh\ and $\Delta\nu_\mathrm{cor,\odot}=135.08\pm0.02$\mh , and $\eta$ is equal to one in case of RGB stars and $1.04\pm0.01$ for RC stars.

\subsection{Seismic fundamental parameters from grid modeling}	\label{sec:gridmodel}

We implement the new nonlinear scalings into the grid-modelling approach from \cite{kal10b}, which compares the observed \num\ and \dcor\ with those of an extensive grid of stellar models where the model seismic parameters are determined from the scaling relations. The grid consists of canonically solar composition-scaled BaSTI isochrones \citep[][, version 5.0.0]{piet2004} with mass-loss parameter $\eta=0.2$. The grid covers more than 4 million models from the sub-giant phase to the asymptotic giant branch (AGB) with initial masses ranging from 0.5 to 4\,M\sun\ and chemical compositions of $(Z,Y) = (0.0003,0.245)$ to $(0.4,0.303)$ with typical increments in $Z$ of 0.001. The Bayesian comparison between observed and model parameters fully accounts for all observational uncertainties including those of the scaling relations, and can use prior information for all model parameter including the evolutionary stage. 

\begin{figure}[t]
	\begin{center}
	\includegraphics[width=0.5\textwidth]{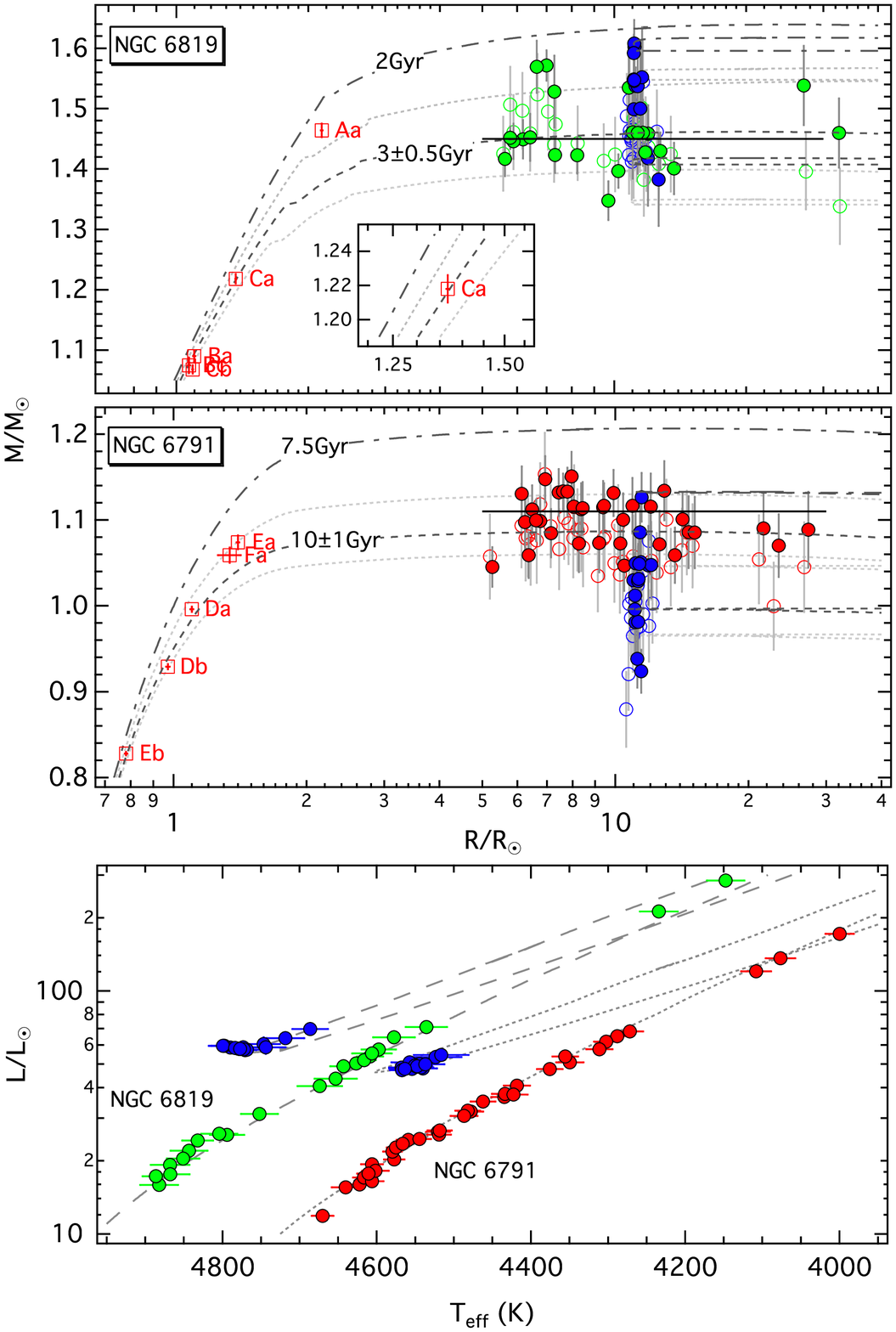}
	\caption{Seismic mass vs. radius from grid modelling (filled circles) and direct computations from scaling relations (open circles) based on \num\ and $\Delta\nu_\mathrm{cor}$ (from Eq.\,\ref{eq:asymrel0}) as seismic input. Symbol colours are the same as in Fig.\,\ref{fig:ClusterSeis}. Also plotted are BaSTI isochrones (dotted, dashed, and dashed-dotted lines) for solar metallicity in the top panel and [Fe/H] = 0.26 in the bottom panel with the numbers indicating the age of the isochrones as well as a number of eclipsing  binaries (red squares), with the labels indicating: Aa $\rightarrow$ WOCS23009 A; Ba, Bb $\rightarrow$ WOCS24009 A \& B; Ca, Cb $\rightarrow$ WOCS40007 A \& B; Da, Db $\rightarrow$ V18 A \& B; Ea, Eb $\rightarrow$ V20 A \& B; Fa $\rightarrow$ V80 A. The inset in the top panel zooms onto the position of the primary component of WOCS40007 (Ca). The bottom panel shows the stars position in the HR-diagram along with a 3 (dotted line) and 10\,Gyr (dashed line) BaSTI isochrone for \na\ and \nb , respectively.} 
	\label{fig:seisMR} 
	\end{center} 
\end{figure}

Unfortunately, our algorithm can only treat stars one by one independently and not as an ensemble with common age and chemical composition. To mimic such an ensemble analysis, we first run our grid-modelling using only \num, \dcor, $T_\mathrm{eff}$, and the evolutionary stage but no prior information on the chemical composition and age of the stars. Even though the resulting best-fit models group around specific metallicities and ages for each cluster the scatter is still quite large. To improve this we then use the median metallicity ($Z=0.033$ for \na\ and $0.020$ for \nb) and age (10.1 and 2.9\,Gyr) found in the first run as priors for a second run, where we use Gaussian priors with a $\sigma$ of 10\%. The resulting best-fit model masses and radii and their uncertainties, are shown in Fig.\,\ref{fig:seisMR}.  As expected the new seismic masses are incompatible with previous estimates from linear scalings and settle at values that are 10-15\% lower than reported by others. 

For \na\ we find the stars to group around $M_\mathrm{RGB}=1.10\pm0.03$ and $M_\mathrm{RC}=1.02\pm0.05$M\sun . The scatter is comparable with the average uncertainties of the individual measurements. Even though our $M_\mathrm{RGB}$ is significantly smaller than found by others, the mass loss of $\Delta M=M_\mathrm{RGB}- M_\mathrm{RC}=0.08\pm0.03$M\sun\ is in perfect agreement with that found by \cite{miglio2012}, who inferred a mass-loss parameter of $\eta=0.1-0.3$. The good agreement is not surprising because the nonlinear effects in our scaling relations in practice cancel out when computing the mass difference of stars with similar \num\ and \dnu. Furthermore we note that the observed mass loss is consistent with that of BaSTI isochrones with $\eta=0.2$ (see Fig.\,\ref{fig:seisMR}).

The situation is less straight forward for \nb, for which we find less good agreement between masses of the individual cluster members. We find an average of $M_\mathrm{RGB}=1.45\pm0.06$M\sun, and a scatter about twice as large as the average uncertainty of the individual measurements. A possible explanation for this extra scatter could be the presence of more than one generation of stars. To verify this is, however, beyond the scope of the present analysis. Also the RC stars scatter about twice as much around the average mass $M_\mathrm{RC}=1.54\pm0.07$M\sun\ than expected from their individual uncertainties. This might be due to the fact that our BaSTI models imply a mass loss between the RGB and RC phase, which leads to overestimation of the mass for RC stars if they do not undergo mass loss in reality. This effect is, however, small ($\sim$2\%) as can be seen from the isochrones in Fig.\,\ref{fig:seisMR} and is therefore within the uncertainties of our analysis.

As a consistency check we also compute $M$ and $R$ directly from the nonlinear scaling relations and find them to generally agree with the values from grid-modelling but with larger uncertainties (see Fig.\,\ref{fig:seisMR}). A summary of the cluster parameters determined from grid-modelling is given in Tab.\,\ref{tab:ClusterStars}. In Fig.\,\ref{fig:seisMR} we also show the stars' positions in the HR-diagram based on the output of our grid-modelling. 

\subsection{Cluster distance moduli}
There are several ways to determine the clusters distance moduli from the seismic observations. One could use the constants $d$ in Eq.\,\ref{eq:numaxfit} and \ref{eq:dnufit}, which represent the mass and true distance modulus terms in the calculated peak frequency (Eq.\,\ref{eq:numax_cal}) and large separation (Eq.\,\ref{eq:dnu_cal}). However, a more direct way is to combine Eq.\,\ref{eq:numax_cal} and \ref{eq:dnu_cal} in order to express the true distance modulus as,
\begin{equation}
DM = 5 \log \nu_\mathrm{max} - 10 \log \Delta\nu + 12.5 \log T_\mathrm{eff} + m_\mathrm{bol} - 4.7554,
\end{equation}
where \num\ and \dnu\ are scaled according to Eq.\,\ref{eq:final_numax} and \ref{eq:final_dnu}, respectively.
From that we find average true distance moduli of $(m-M)_0=13.11\pm0.03$ (\na) and $11.91\pm0.03$\,mag (\nb), where the uncertainties represent the spread of the individual values (and not the actual error of the mean, which would be about five times smaller). Our distance moduli are in agreement with, but more precise than, previous seismic and non-seismic determinations and translate into distances of $4.19\pm0.06$\,kpc (\na) and $2.41\pm0.03$\,kpc (\nb).

\begin{table*}[t]
\begin{small}
\begin{center}
\caption{RGB mass, mass loss, true distance modulus, and age for both clusters from selected sources in the literature and determined in the present analysis (based on the nonlinear scalings with \num\ and $\Delta\nu_\mathrm{cor}$ as seimic input).
\label{tab:ClusterStars}}
\begin{tabular}{r|r|cccc|ccc}
\hline\hline
\noalign{\smallskip}
\multicolumn{2}{c}{}&\multicolumn{4}{|c|}{\na} & \multicolumn{3}{c}{\nb} \\
\noalign{\smallskip}
\hline
\noalign{\smallskip}
\multicolumn{2}{c|}{}&$M_\mathrm{RGB}$ & $\Delta M$  & $(m-M)_0$ & age &$M_\mathrm{RGB}$ & $(m-M)_0$ & age\\
\multicolumn{2}{c|}{}&\multicolumn{2}{c}{[M\sun]} & [mag] & [Gyr] & [M\sun] & [mag] & [Gyr]\\
\noalign{\smallskip}
\hline
\noalign{\smallskip}
\cite{basu2011} &seismology& $1.20\pm0.01$ & ... & $13.11\pm0.06$ & 6.8$-$8.6 & $1.68\pm0.03$ & $11.85\pm0.05$ & 2$-$2.4\\
\cite{miglio2012} &seismology& $1.23\pm0.01$ & $0.09\pm0.05$ & ... & ... & $1.61\pm0.02$ & ... & ...\\
\cite{wu2014b,wu2014b} &seismology& $1.25\pm0.03$  & ... & $13.09\pm0.10$ & ... & $1.75\pm0.05$ & $11.88\pm0.14$  & ...\\
\cite{handberg2017} &seismology& ...  & ... & ... & ... & $1.61\pm0.02$ & ...  & ...\\
\cite{An2015} &isochrones& ... & ... & $13.04\pm0.08$ & $9.5\pm0.3$ & ... & ... & ...\\
\cite{Balona2013} &isochrones& ... & ... & ... & ... & ... & $11.94\pm0.04$ & 2.5\\
\cite{Brogaard2012} & binaries & $1.15\pm0.02$ & ... & ... & $8.3\pm0.5$ & ... & ... & ... \\
\cite{Grundahl2008} & binaries & ... & ... & ... & 7.7$-$9 & ... & ... & ... \\
\cite{Jeffries2013} & binaries & ... & ... & ... & ... & ... & ... & $2.9-3.7$\\
\cite{Brewer2016} & binaries & ... & ... & ... & ... & ... & ... & $2.4\pm0.2$\\
\noalign{\smallskip}
this work &seismology& $1.10\pm0.03$ & $0.08\pm0.03$ & $13.11\pm0.03$ & $10.1\pm0.9$ & $1.45\pm0.06$ & $11.91\pm0.03$ & $2.9\pm0.3$\\
\hline
\end{tabular}
\end{center}
\end{small}
\end{table*}

\subsection{Cluster ages}
A direct consequence of the fact that the classic seismic scaling relations overestimate the mass of red giants is that they significantly underestimate their ages. With our nonlinear scaling relations implemented in the grid-modelling we measure the ages to be $10.1\pm0.9$ for \na\ and $2.9\pm0.3$\,Gyr for \nb, which is about 30\% older than reported from previous seismic analyses. In principle, any age determination suffer from various sources of uncertainties. For example, the isochrone fiting to an observed CMD that are frequently used to estimate cluster ages suffer from the usual difficulties of measuring interstellar reddening/extinction and the transformation of the model parameters into observable colours. A more robust and direct method is to compare the mass and radius of isochrones to those of EB cluster members. 

There are a number of known EB members in \na. \cite{Grundahl2008} carried out a detailed analysis of the system V20 with one component located close to the cluster turnoff. This allowed them to measure the cluster age to better than a few per cent within a given set of isochrones. They did, however, also find that the actual age largely depends on the chosen set of isochrones, and hence reported an age ranging from 7.7 to 9\,Gyr, with the latter estimate coming from BaSTI isochrones. The reason for this ambiguity might be due to the high metallicity of the cluster or due to the role convective overshoot plays in the morphology of isochrones near the turnoff. \cite{Brogaard2011,Brogaard2012} used two additional binary systems (V18 and V80). They find that isochrones that are required to fit the EBs in the mass-radius diagram as well as in the clusters colour-magnitude diagram give a best fit (within their set of Victoria isochrones) for an age of $8.3\pm0.5$\,Gyr, which corresponds to a mass on the lower RGB of $1.15\pm0.02$\,M\sun . \cite{Brogaard2016} further measured the mass of the giant component of the EB V9 to be $1.14\pm0.02$\,M\sun , which is again more compatible with our $M_\mathrm{RGB}$ than with that from previous seismic analyses.

We do not perform isochrone fitting but exemplarily show isochrones in Fig.\,\ref{fig:seisMR} that correspond to the result of our grid modelling. In fact, the EBs are more consistent with a $10\pm1$\,Gyr age-range than with previous seismic age estimates and therefore support the low RGB mass.

\nb\ also includes a number of known detached eclipsing binary systems \citep[e.g.,][]{Jeffries2013,Sandquist2013,Brewer2016}, most notably WOCS 24009 and WOCS 40007, whose components are all located at the upper main sequence of the cluster ($1-1.25$M\sun; see Fig.\,\ref{fig:seisMR}). The most massive and therefore most age sensitive star amongst them is WOCS 40007 A. \cite{Jeffries2013} determined an age of $3.1\pm0.4$\,Gyr from isochrone fitting to the well-defined mass and radius of the star. Even though this estimate is affected by systematic uncertainties it is quite clear that the real age of the star exceeds the commonly adopted age of the cluster of $2-2.5$\,Gyr. A potentially even more age sensitive star is WOCS 23009 A, which is located near the clusters' turnoff. \cite{Brewer2016} used all measured EBs to determine a cluster age of $2.4\pm0.2$\,Gyr. However, this result relies, mostly on the mass of WOCS 23009 A, which might be questioned because the very low-mass companion hampers a dynamic mass estimate. Instead, its mass was determined from the isochrones used to constrain the cluster age. We argue that this represents a circular argument, so that the star should be excluded from the analysis. As can be seen from Fig.\,\ref{fig:seisMR}, the remaining binaries are in favour of the older isochrones and therefore support the low RGB mass of about 1.45\,M\sun\ instead of the previously reported 1.61--1.75\,M\sun. 

Following the findings of \cite{Grundahl2008} we note that it might well be that our cluster ages from BaSTI isochrones mark more an upper limit than an accurate determination. In this context we note that \cite{Stancliffe2016} have shown that solar evolutionary tracks from various stellar evolution codes differ by up to 150\,K when leaving the main sequence at significantly different ages, with BaSTI providing the oldest models. It is therefore not surprising that the two clusters ages are still a matter of debate. However, to solve this is beyond the scope of the present analysis. 

\begin{figure}[t]
	\begin{center}
	\includegraphics[width=0.5\textwidth]{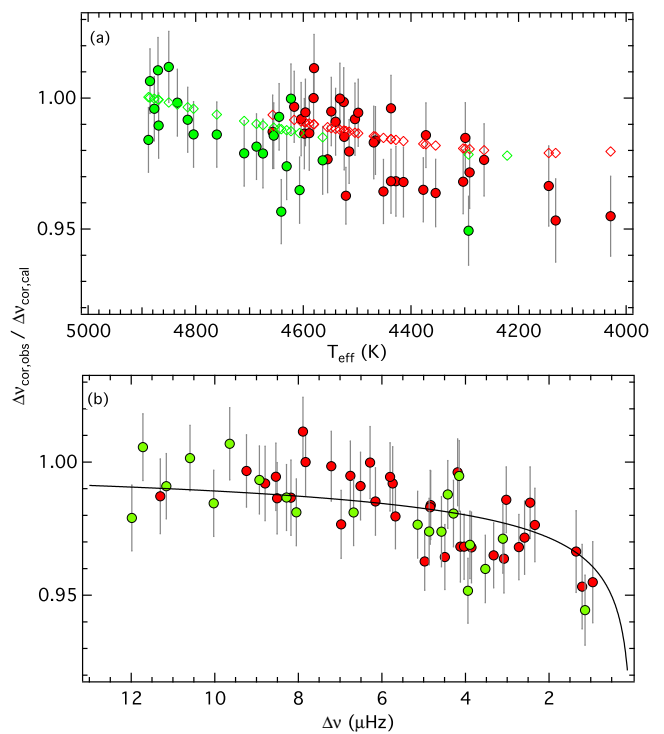}
	\caption{Ratio of observed and calculated (see Eq.\,\ref{eq:dnu_cal}) large separation for RGB stars in \na\ (red circles) and \nb\ (green circles). Panel (a) shows this ratio as a function of the effective temperatur along with the expected values from \cite{guggenberger2016} (diamond symbols). Panel (b) shows the ratio as function of \dnu\ along with nonlinear scaling of the solar reference (solid line), i.e., the term in square brackets of Eq.\,\ref{eq:final_dnu}.} 
	\label{fig:teffdnu} 
	\end{center} 
\end{figure}

\subsection{Metallicity dependence}

With the true distance modulus and mass available we can directly compute $\Delta\nu_\mathrm{cor,cal}$ from Eq.\,\ref{eq:dnu_cal} using the linear scaling relations, and compare it to the observed values. This is shown in Fig.\,\ref{fig:teffdnu}a, where we plot the ratio between the observed and calculated large separation as a function of effective temperature.  As suggested from models by \citet{white2011}, we do indeed find a trend that the scaled large separation increasingly overestimates the observations towards lower effective temperature on the RGB. However, our uncertainties are too large to reveal a potential dependence of this trend on metallicity as suggested by the models. We can therefore neither confirm nor disprove any metallicity influence on the large separation. We do, however, show in Fig.\,\ref{fig:teffdnu}b that this trend is well reproduced by the nonlinear $\Delta\nu - \bar\rho$ scaling, without the need to incorporate the stellar metallicity. 

\section{Implications for the APOKASC sample}
\begin{figure}[t]
	\begin{center}
	\includegraphics[width=0.5\textwidth]{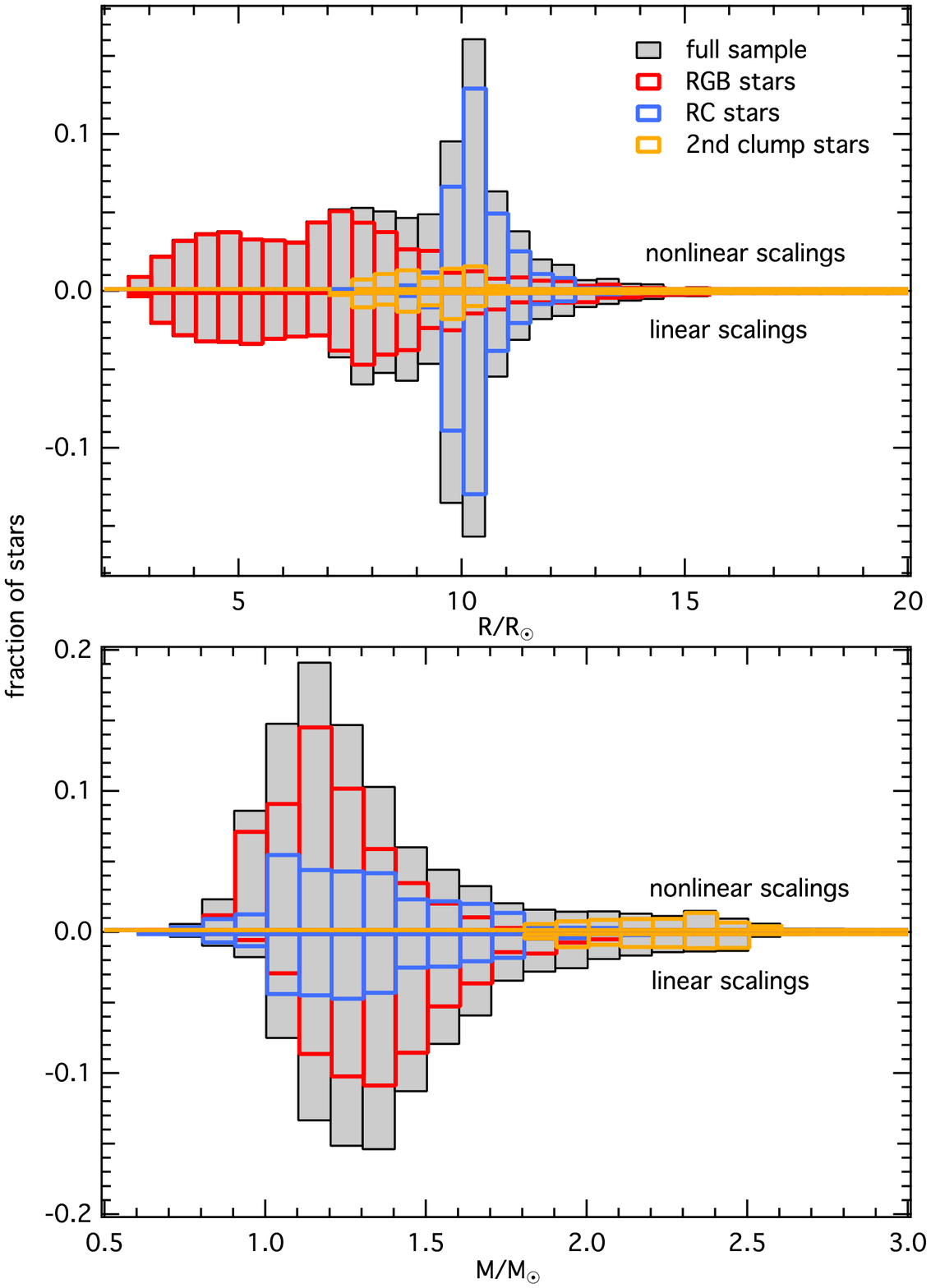}
	\caption{Radius (top panel) and mass (bottom panel) distributions of about 5100 red giants from the APOKASC sample computed from grid modelling based on our new nonlinear scalings with \num\ and $\Delta\nu_\mathrm{cor}$ as seismic input (upwards pointing) and the classical linear scalings (downwards pointing). The full sample and subsamples corresponding to the different evolutionary stages are shown by grey and coloured bars, respectively.} 
	\label{fig:RMhisto} 
	\end{center} 
\end{figure}

Apart from the consequences for the cluster ages and RGB masses shown above, our new scaling relations will also impact the seismic fundamental parameter determination of other red giants. A prominent set of red giants that is currently vigorously investigated is the so-called APOKASC sample \citep[e.g.,][]{Elsworth2017b}. APOKASC is a collaboration between APOGEEs' spectroscopic survey in the near IR at Apache Point (SDSS) and the Kepler Asteroseismic Consortium (KASC) for the study of cool stars in the \kep\ field with the aim to take advantage of the high-quality spectroscopic and asteroseismic data to derive precise fundamental parameters for a large number of stars. The resulting catalog includes about 6600 red giant stars \citep{pin2014}, which are spectroscopically and asteroseismically characterised. Among those, we can determine the curvature and glitch corrected large separation $\Delta\nu_\mathrm{cor}$ (see Sec.\,\ref{sec:asymDnu}) with sufficient accuracy (the median uncertainty is about 0.13\%) from the radial mode spectrum of about 5100 stars. Details of the mode fitting and the curvature and glitch analysis are subject to a separate paper (Kallinger et al., in preparation) and first results about the associated mode amplitudes and line widths are given by \cite{vrard2018}. 

We then use the grid-modelling approach described in Sec.\,\ref{sec:gridmodel} to determine the seismic masses and radii from the \num , \dnu$_\mathrm{cor}$, $T_\mathrm{eff}$, and evolutionary stage measurements of these stars. To study the implications of the nonlinear  scalings on the mass and radius distributions of red giants in the \kep\ field, we computed the seismic masses and radii twice: once based on the original seismic scalings and once based on our new nonlinear scalings. The results are shown in Fig.\,\ref{fig:RMhisto}. While the radius distribution is basically unaffected, we find a significant shift towards lower masses in the mass distribution. Interestingly, this shift is largely caused by the RGB stars in the sample. Whereas the distribution of core-He burning stars (red clump and secondary clump stars) is relatively unaffected, the peak in the distribution of RGB stars shifts from the 1.3$-$1.4\,M\sun\ bin to the 1.1$-$1.2\,M\sun\ bin, when implementing the nonlinear seismic scalings instead of the original linear scalings in the grid modelling. Numerical tests show that the reason core-He burning stars are basically unaffected is because the ``nonlinear effect'' in the $\Delta\nu -$scaling is partly compensated by the correction factor $\eta$ (which is different from one for core-He burning stars). 

The about 15\% shift in mass for RGB stars cannot be explained by the uncertainties of the individual measurements, which are on the order of 5\%. It is therefore a real effect and should be of particular interest for future galactic archeology studies \citep[such as][]{Anders2017,Stello2017}, which attempt to use seismic mass and radius distributions of different parts of the Milkyway to study the structural and chemical evolution of our Galaxy.
 
\section{Summary}

In recent years the seismic scaling relations for \num\ and \dnu\ have become increasing important in various fields of astrophysics because they can be used to estimate the mass and radius of stars showing solar-like oscillations. With the high-precision photometry of \textit{Kepler}, the uncertainties in the seismic observables are now small enough to estimate masses and radii with a precision of only a few per cent. Even though this seems to work quite well for main-sequence stars, there is empirical evidence, mainly from studies of eclipsing binary systems, that the seismic scaling relations systematically overestimate the mass and radius of red giants by about 5 and 15\%, respectively. 

Various model-based corrections to the $\Delta\nu -$scaling reduce the problem but do not solve it. Furthermore, such corrections suffer from systematic errors in the models, which clearly affect the models' large frequency separation but which are not accounted for in the corrections. It is therefore not yet clear if  these corrections are of any physical relevance for the $\Delta\nu -$scaling or simply the consequence of the systematic errors in the models. 

Our goal was therefore to revise the seismic scaling relations in order to account for the known systematic mass and radius discrepancies in a completely model-independent way. This can be done best by comparing the dynamic and seismic fundamental parameters of eclipsing binaries hosting at least one star with detectable solar-like oscillations. So far there are six such systems known, with measurements accurate enough for a conclusive analysis. To minimise potential biases in the dynamic parameters we used the results from three different groups \citep{gaulme2016,Brogaard2017,themessl2016} based on different instrumentation and analysis methods. We used probabilistic methods to re-examine the global oscillation parameters of the giants in the binary systems. From this we determine the corresponding seismic fundamental parameters and find them to agree with the dynamic parameters if we adopt nonlinear seismic scaling relations. Even though the sample is quite small the Bayesian evidence is clearly in favour of the new nonlinear scalings. We also find that the results are best when using a curvature and glitch corrected large separation as input for the $\Delta\nu -$scaling. 

We then compare the seismic parameters of about 60 red giants observed by \kep\ in two open clusters to those scaled from independent measurements and find the same nonlinear behaviour as for the eclipsing binaries. Our final proposed scaling relations (Eq.\,\ref{eq:final_numax} and \ref{eq:final_dnu}) are based on both samples (eclipsing binaries and cluster stars) and cover a broad range of evolutionary stages from red giant branch stars to core-helium burning red clump stars.

It has been suggested that at least the $\Delta\nu -$scaling should in some sense be sensitive to stellar metallicity. We find that our samples are too small and/or not accurate enough to proof or disproof such a metallicity dependency. On the other hand, we could show that a previously claimed ``metallicity effect'' that was made responsible for the discrepancy between seismic and independent estimates of the mass of some highly evolved (luminous) halo and thick-disk stars can be equally well explained with our (metallicity-independent) nonlinear scalings. The difference between the previous linear and our new nonlinear relations increases towards more luminous stars. 

A direct consequence of the new scaling relations is that the average mass of stars on the ascending giant branch in both investigated clusters reduces on average by about 12\%, allowing us to revise the clusters' true distance moduli and to conclude that both clusters are older than suggested by previous seismic investigations. We further note that independent estimates of the masses and radii of several eclipsing binaries located before or close to the turnoff in both clusters are more consistent with our new $M_\mathrm{RGB}$ measurements than with those based on linear scaling relations. Apart from its scientific value we interpret this as another indication that our revised scaling relations provide more realistic estimates of the mass and radius of a red giant than the classical relations. 

We also checked how the revised scalings affect the mass and radius distributions of red giants in the full \kep\ field, which is an input for galactic archeology studies. We find that while the radius distribution is largely unaffected, the mass distribution changes significantly, which is predominantly caused by an about 15\% reduction of the mass distribution of RGB stars.

The revised seismic scaling relations presented here are most likely still a simplification of the intrinsic relation between the seismic quantities and the fundamental parameters of a star. To check and potentially improve this would require large samples of stars with accurate seismic and independent measurements of their fundamental parameters, which are currently not available. The original \kep\ field does not include many more eclipsing binaries, which are sufficient for such an analysis. Observations with the K2 mission \citep{Howell2014} might reveal new systems but their seismic observables are not accurate enough. We might expect some improvements \citep[in the sense of][]{Huber2017} from future \textit{Gaia} data releases for stars in the original \kep\ field but these stars lack independent mass constraints. More promising in the near future are the observations with one-year coverage near the ecliptic poles by TESS \citep{Ricker2014}, which will very likely reveal many new eclipsing binary systems suitable for further testing and improving the seismic scaling relations.    
    
Finally, we advise the reader to some caution when using the scaling relations presented here. They are calibrated for well-defined seismic input parameters and any deviations from these definitions will cause systematic uncertainties  that can easily exceed the random uncertainties. For the peak frequency, one has to consider an appropriate treatment of the granulation background. For the large frequency separation we note that a curvature and glitch corrected global value should be preferred. If this is not accessible, we also provide a calibration for a local (central) value of \dnu . For other definitions of \num\ and \dnu\ one needs to re-calibrated the corresponding scalings to achieve the best possible precision and accuracy.

\begin{acknowledgements}
The authors gratefully acknowledge the \textit{Kepler Science Team} and all those who have contributed to making the \textit{Kepler} mission possible. Funding for the \textit{Kepler Discovery mission} is provided by NASA's Science Mission Directorate. This publication makes use of data products from the Two Micron All Sky Survey, which is a joint project of the University of Massachusetts and the Infrared Processing and Analysis Center/California Institute of Technology, funded by the National Aeronautics and Space Administration and the
National Science Foundation. TK is grateful for funding via the Austrian Space Application Programme (ASAP) of the Austrian Research Promotion Agency (FFG) and BMVIT. PGB acknowledges the support of the Spanish Ministry of Economy and Competitiveness (MINECO) under the programme 'Juan de la Cierva incorporacion' (IJCI-2015-26034). DS is the recipient of an Australian Research Council Future Fellowship (project number FT1400147). RAG acknowledges the support of the CNES-PLATO grant.
\end{acknowledgements}


\bibliographystyle{aa}
\bibliography{32831}

\end{document}